\documentclass[aps,prb,twocolumn,superscriptaddress,floatfix,longbibliography]{revtex4-2}

\usepackage{lineno}
\usepackage{amsmath,amssymb} 
\usepackage{bm} 
\usepackage{graphicx} 
\usepackage{comment} 
\usepackage{textcomp} 
\usepackage{color,soul}

\usepackage[colorinlistoftodos]{todonotes}
\usepackage{siunitx}
\usepackage{enumitem}
\setlist{noitemsep,leftmargin=*,topsep=0pt,parsep=0pt}
\usepackage{flushend}
\usepackage{upgreek}

\usepackage{xr}
\makeatletter

\newcommand*{\addFileDependency}[1]{
\typeout{(#1)}
\@addtofilelist{#1}
\IfFileExists{#1}{}{\typeout{No file #1.}}
}\makeatother

\usepackage{xcolor} 
\definecolor{lightgray}{gray}{0.6}
\definecolor{medgray}{gray}{0.4}

\usepackage{hyperref}
\hypersetup{
bookmarks=true,         
unicode=false,          
pdftoolbar=true,        
pdfmenubar=true,        
pdffitwindow=false,     
pdftitle={Certificate},    
pdfauthor={Dr. Harish Kumar},     
pdfsubject={TEQIP certificates},   
pdfcreator={Dr. Harish Kumar},   
pdfproducer={},  
pdfkeywords={Certificates,} {TEQIP} {Participation}, 
pdfnewwindow=true,      
colorlinks=true,
urlcolor= blue,
citecolor=blue,
linkcolor= blue,
}

\newif\ifptitle
\newif\ifpnumber
\newcounter{para}

\ptitletrue  
\pnumbertrue  

\newcommand{\mytitle}{Driven bright solitons on a mid-infrared laser chip}

\begin{document}
\title{\mytitle}

\author{Dmitry Kazakov}
\email[]{kazakov@seas.harvard.edu}
\affiliation{Harvard John A. Paulson School of Engineering and Applied Sciences, Harvard University, Cambridge, MA 02138, USA}

\author{Theodore P. Letsou}
\affiliation{Harvard John A. Paulson School of Engineering and Applied Sciences, Harvard University, Cambridge, MA 02138, USA}
\affiliation{Department of Electrical Engineering and Computer Science, Massachusetts Institute of Technology, Cambridge, MA 02142, USA}

\author{Marco Piccardo}
\affiliation{Harvard John A. Paulson School of Engineering and Applied Sciences, Harvard University, Cambridge, MA 02138, USA}
\affiliation{Department of Physics, Instituto Superior Técnico, Universidade de Lisboa, 1049-001 Lisbon, Portugal}
\affiliation{Instituto de Engenharia de Sistemas e Computadores – Microsistemas e Nanotecnologias (INESC MN), 1000-029 Lisbon, Portugal}

\author{Lorenzo L. Columbo}
\affiliation{Dipartimento di Elettronica e Telecomunicazioni, Politecnico di Torino, 10129 Torino, Italy}

\author{Massimo Brambilla}
\affiliation{Dipartimento di Fisica Interateneo, Università e Politecnico di Bari, 70126 Bari, Italy}
\affiliation{CNR-Istituto di Fotonica e Nanotecnologie, 70126 Bari, Italy}

\author{Franco Prati}
\affiliation{Dipartimento di Scienza e Alta Tecnologia, Università dell’Insubria, 22100 Como, Italy}

\author{Sandro Dal Cin}
\affiliation{Institute of Solid State Electronics, TU Wien, 1040 Vienna, Austria}

\author{Maximilian Beiser}
\affiliation{Institute of Solid State Electronics, TU Wien, 1040 Vienna, Austria}

\author{Nikola Opa{\v{c}}ak}
\affiliation{Institute of Solid State Electronics, TU Wien, 1040 Vienna, Austria}

\author{Pawan Ratra}
\affiliation{Harvard John A. Paulson School of Engineering and Applied Sciences, Harvard University, Cambridge, MA 02138, USA}
\affiliation{Department of Electrical and Electronic Engineering, Imperial College London, London SW7 2BX, United Kingdom}

\author{Michael Pushkarsky}
\affiliation{DRS Daylight Solutions, San Diego, CA 92127, USA}

\author{David Caffey}
\affiliation{DRS Daylight Solutions, San Diego, CA 92127, USA}

\author{Timothy Day}
\affiliation{DRS Daylight Solutions, San Diego, CA 92127, USA}

\author{Luigi A. Lugiato}
\affiliation{Dipartimento di Scienza e Alta Tecnologia, Università dell’Insubria, 22100 Como, Italy}

\author{Benedikt Schwarz}
\affiliation{Harvard John A. Paulson School of Engineering and Applied Sciences, Harvard University, Cambridge, MA 02138, USA}
\affiliation{Institute of Solid State Electronics, TU Wien, 1040 Vienna, Austria}

\author{Federico Capasso}
\email[]{capasso@seas.harvard.edu}
\affiliation{Harvard John A. Paulson School of Engineering and Applied Sciences, Harvard University, Cambridge, MA 02138, USA}

\date{\today}


\begin{abstract}
    Despite the ongoing progress in integrated optical frequency comb technology~\cite{Chang2022IntegratedTechnologies}, compact sources of short bright pulses in the mid-infrared wavelength range from 3~$\upmu$m to 12~$\upmu$m so far remained beyond reach. The state-of-the-art ultrafast pulse emitters in the mid-infrared are complex, bulky, and inefficient systems based on the downconversion of near-infrared or visible pulsed laser sources. Here we show a purely DC-driven semiconductor laser chip that generates one picosecond solitons at the center wavelength of 8.3~$\upmu$m at GHz repetition rates. The soliton generation scheme is akin to that of passive nonlinear Kerr resonators~\cite{Kippenberg2018DissipativeMicroresonators}. It relies on a fast bistability in active nonlinear laser resonators, unlike traditional passive mode-locking which relies on saturable absorbers~\cite{Liu2019High-channel-countCapacity} or active mode-locking by gain modulation in semiconductor lasers~\cite{Hillbrand2019PicosecondLaser}. Monolithic integration of all components --- drive laser, active ring resonator, coupler, and pump filter --- enables turnkey generation of bright solitons that remain robust for hours of continuous operation without active stabilization. Such devices can be readily produced at industrial laser foundries using standard fabrication protocols. Our work unifies the physics of active and passive microresonator frequency combs, while simultaneously establishing a technology for nonlinear integrated photonics in the mid-infrared~\cite{Ren2023IntegratedRoadmap}.

    

    


\end{abstract}

\maketitle

Short optical pulses have revolutionized a range of applications, from high-resolution imaging and ultrafast spectroscopy to optical communications~\cite{Marin-Palomo2019Comb-basedDiode}, laser-based medical procedures~\cite{Dausinger2004FemtosecondApplications}, and light ranging~\cite{Riemensberger2020MassivelyMicrocomb}. They play a pivotal role in nonlinear optics, enabling a variety of phenomena such as supercontinuum generation~\cite{Dudley2006SupercontinuumFiber} and optical frequency conversion that are at the heart of optical atomic clocks~\cite{Takamoto2020TestClocks}. Over the past two decades, there has been a remarkable progress in miniaturizing pulsed optical sources and transitioning them from tabletop experimental setups to compact photonic integrated chips. The two primary technologies driving this effort are semiconductor mode-locked lasers (SMLLs) and nonlinear microresonator frequency combs~\cite{Chang2022IntegratedTechnologies}. These technologies may reduce the size and complexity of optical atomic clocks~\cite{Drake2019Terahertz-RateClockwork,Spencer2018AnPhotonics}, pave the way to terabit-per-second telecommunication links~\cite{Marin-Palomo2017Microresonator-basedCommunications,Liu2019High-channel-countCapacity}, and provide new tools for linear and nonlinear absorption spectroscopy~\cite{Cundiff2013OpticalSpectroscopy,Picque2019FrequencySpectroscopy}. The near-infrared wavelength range ($0.8-2.5$ $\upmu$m) has an abundance of compact pulsed sources based both on SMLLs and nonlinear resonators. In contrast, there has been a scarce number of demonstrations of integrated photonic pulse generators in the mid-infrared wavelength range ($3-12$ $\upmu$m), strategic to applications in gas sensing and spectroscopy. The only devices in this range that are compatible with photonic integration and capable of directly generating coherent mid-infrared radiation are interband cascade lasers (ICLs) and quantum cascade lasers (QCLs). 

While recently there have been significant advancements in pulse generation using these platforms, major limitations still exist that hinder their deployment in applications. ICLs, that operate at $3-6$ $\upmu$m, have shown modest milliwatt-level average optical power and the pulse durations limited to several picoseconds~\cite{Hillbrand2019PicosecondLaser}. QCLs, which are most efficient in the $4.5 - 12$ $\upmu$m range, now routinely reach Watt-level output power~\cite{Wang2020ContinuousOperation,Schwarz2017Watt-LevelLaser/Detector} and have recently entered the femtosecond regime~\cite{Taschler2021FemtosecondLaser}. However, achieving subpicosecond pulse duration required spectral broadening and external pulse shaping involving strong radiofrequency modulation of the laser bias, a tabletop delay line, and an optical isolator to prevent laser destabilization due to optical feedback. These aspects pose a challenge for miniaturization of the system. 

\begin{figure*}[!t]
    \includegraphics[clip=true,width=\textwidth]{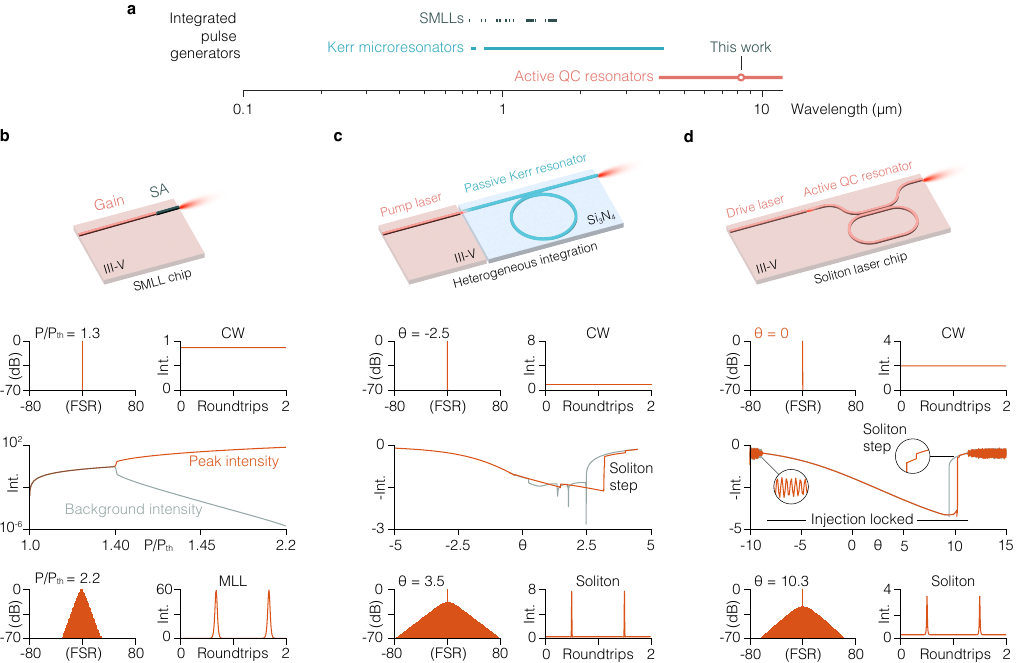}
    \caption{\textbf{Pulse generation in integrated resonator systems.} \textbf{a}. Spectral coverage of the three photonic integrated technology stacks for on-chip short pulse generation. Semiconductor mode-locked laser (SMLL) and Kerr microresonator spectral coverage data are taken from ref.~\cite{Chang2022IntegratedTechnologies}. \textbf{b}. SMLL comprises a forward biased gain section and a reverse biased saturable absorber (SA) section. Reverse biasing the SA section ($\mathrm{P}_{\mathrm{SA}} = -0.8\mathrm{P}_{\mathrm{th}}$) and increasing the pumping of the forward-biased gain section allow a transition from a CW waveform to a mode-locked pulsed operation. The panels show the simulated waveforms and optical spectra of the intracavity field in a SMLL at different values of the gain section pumping. \textbf{c}. Passive Kerr resonator optically pumped by an external laser experiences a bistable behavior in the output intensity as a function of the detuning $\uptheta$ of the laser wavelength from the Kerr cavity resonance. Here the middle panel shows the mean intracavity intensity as function of detuning. The intensity values are artificially flipped by multiplying with $-1$ to facilitate the visual comparison with the experimental plots later in the manuscript, where not the intracavity intensity, but the measured output intensity is shown. On a forward scan (blue-detuned to red-detuned, orange curve in the middle panel) the resonator transmission shows a series of steps signifying generation of one or several cavity solitons, which cannot be generated on a backward scan (gray curve in the middle panel). The simulations are based on the GLLE~\cite{Columbo2021UnifyingLasers}. State-of-the-art fabrication technology now allows monolithic integration of the pump laser with a passive resonator~\cite{Xiang20233DPhotonics}. \textbf{d}. Active quantum cascade (QC) resonator with an on-chip integrated drive laser. The soliton excitation scheme is similar to that of passive Kerr resonators and entails scanning the frequency of the drive laser through the lasing resonance of the active racetrack resonator. The associated output intensity (orange curve --- forward scan, gray curve --- backward scan) exhibits optical bistability and a characteristic step in transmission when generating one or several cavity solitons on a forward wavelength scan.}
     \label{fig:pulse_generators}
\end{figure*}

We address this challenge by introducing a new feedback-insensitive and isolator-free semiconductor laser architecture, capable of generating bright picosecond, mid-infrared pulses directly on a laser chip without any external compression. Pulse generation is enabled by a new driving scheme that does not rely on active or passive mode-locking as in SMLLs; instead, it is inspired by the soliton excitation techniques developed for passive Kerr microresonators and is enabled by an experimental unification of the physics of dissipative solitons in passive and active cavities based on a generalized Lugiato-Lefever equation (GLLE)~\cite{Columbo2021UnifyingLasers}. This equation extends the LLE --- which was originally formulated for resonators without population inversion~\cite{Lugiato1987SpatialSystems} --- to ring lasers driven above threshold~\cite{Lugiato1988CooperativeLasers,Piccardo2020FrequencyTurbulence}. Although we perform our demonstration using a QCL emitting at $8.3$ $\upmu$m, this approach applies to any device of such class across the mid-infrared. The chip-scale pulse generators shown here thus extend the spectral gamut covered by the SMLLs and Kerr microresonators to the entire mid-infrared range (Fig.~\ref{fig:pulse_generators}\textbf{a}).   

\begin{figure*}[!t]
    \includegraphics[clip=true,width=\textwidth]{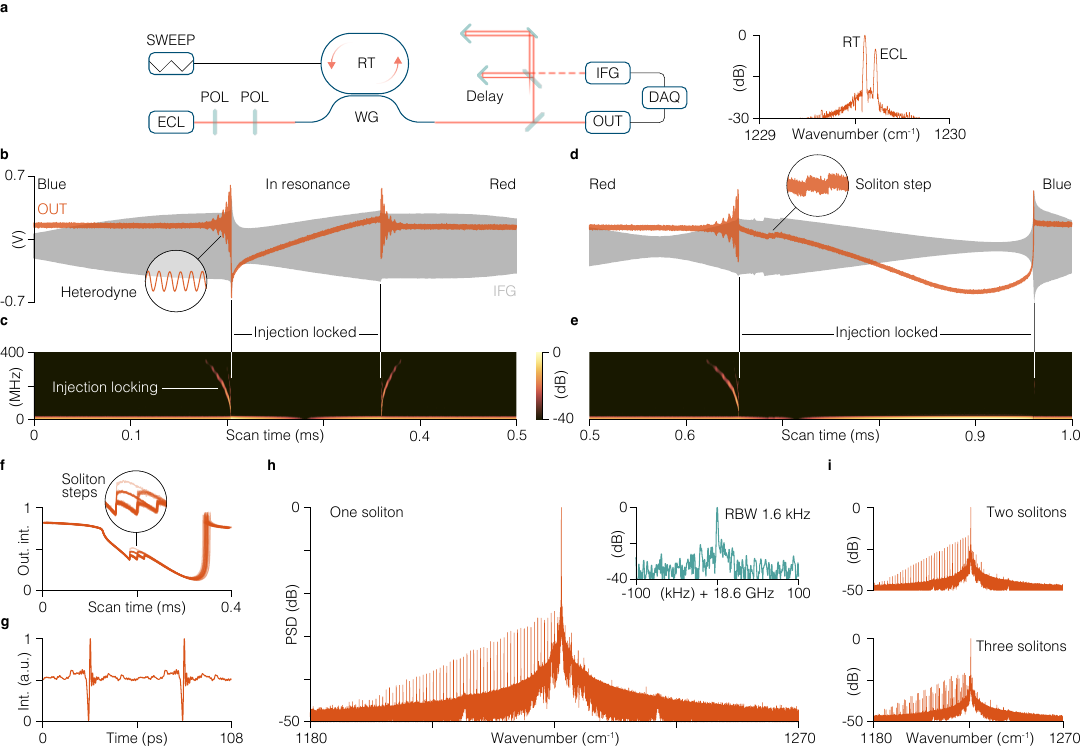}
    \caption{\textbf{Driven bright solitons in an active resonator.} This demonstration relies on an off-chip external drive laser. \textbf{a}. Experimental setup for external soliton driving. SWEEP, function generator used to apply triangular sawtooth modulation to the integrated heater (see Methods for further details on resonance tuning). ECL, external cavity laser. POL, polarizer. RT, racetrack resonator. WG, waveguide coupler. IFG, interferogram signal. OUT, output intensity detector. DAQ, data acquisition board. The high-resolution optical spectrum shows RT and ECL in proximity to each other (ECL on the blue side of the RT) at the onset of the detuning scan. \textbf{b}. Output intensity (orange) and interferogram (gray) acquired upon a forward wavelength scan (blue-detuned to red-detuned). \textbf{c}. Short-time FFT of the output intensity trace on the forward scan showing the heterodyne notes at the onset of injection locking of the RT lasing frequency to that of the ECL. \textbf{d}. Output intensity and interferogram on a backward wavelength scan (red-detuned to blue-detuned). \textbf{e}. Short-time FFT of the output intensity trace on the backward scan showing the characteristic step signifying soliton generation. \textbf{f}. Overlaid output intensity curves over one thousand backward scans, showing multistability of the soliton states. The traces are aligned so that the first steps in the output intensity coincide. \textbf{g}. Reconstructed output temporal waveform of the driven soliton generator over two consecutive roundtrips. \textbf{h}. Optical spectrum of the waveform in \textbf{g}. The inset shows the corresponding RF spectrum of the intermode beatnote. \textbf{i}. Optical spectra of multiple soliton states, obtained on the backward scan, corresponding to the different steps in the output intensity seen in \textbf{f}.}
     \label{fig:forward_backward_exp}
\end{figure*}

\begin{figure*}[!t]
    \includegraphics[clip=true,width=\textwidth]{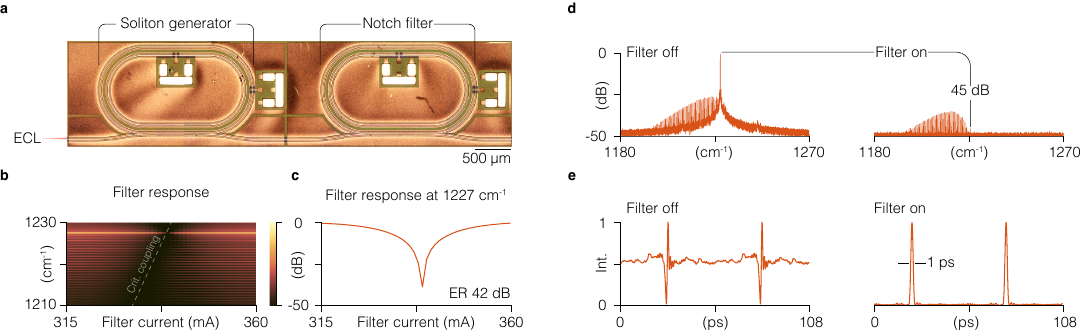}
    \caption{\textbf{On-chip pump filtering.} \textbf{a}. Optical microscope image of the active resonator photonic integrated chip. The chip contains two identical racetrack resonators connected via a bus waveguide --- one for soliton generation, another one --- for pump filtering. Integrated heaters are implemented as concentric racetracks on the inner side of the soliton generator and of the filter. \textbf{b}. Spectrogram of the soliton state from the left racetrack resonator in \textbf{a}, as the filter response is tuned with the pump current. Critically coupled ring resonator progressively filters out comb teeth one by one, shown by the dashed line. \textbf{c}. The filter response at the pump mode wavelength (1227 cm$^{-1}$), showing the extinction ratio (ER) of 42 dB. \textbf{d}. Optical spectra of the soliton generator when the filter is switched off and switched on. The suppression of the pump frequency by 45 dB is achieved by fine-tuning the current of the filter and of the integrated heater. \textbf{e}. Reconstructed temporal waveforms of the soliton generator over two consecutive cavity roundtrips with filter off and filter on.}
     \label{fig:integrated_filter}
\end{figure*}

\begin{figure*}[!t]
    \includegraphics[clip=true,width=\textwidth]{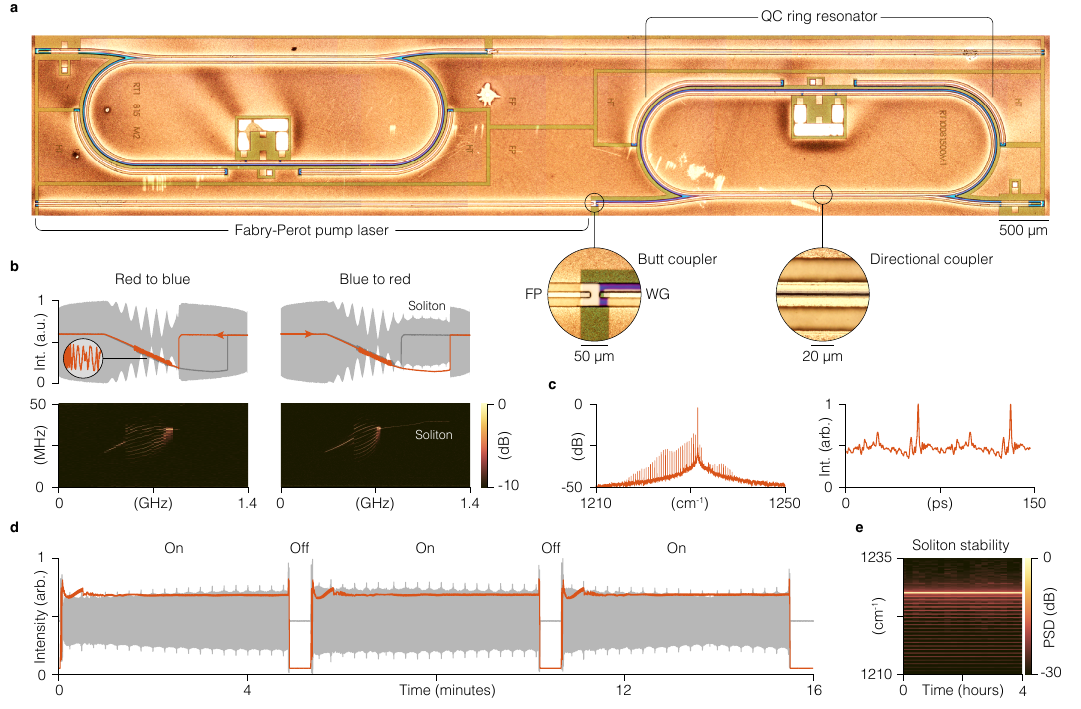}
    \caption{\textbf{Integrated turnkey soliton generator.} \textbf{a}. Optical microscope image of the QC photonic integrated chip. Shown chip contains two identical devices each comprised by four components: a Fabry-Perot drive laser (FP), a waveguide coupler (WG), a resistive heater (HT) and a racetrack resonator (RT). Insets show blown up micrographs of the coupling regions between the FP and the WG (butt coupler) and between the WG and the RT (directional coupler). \textbf{b}. Top: output intensity (orange), interferogram (gray), and the optical spectra of the RT resonator on a backward and on a forward scan of the FP through the RT lasing resonance. Bottom: spectrogram of the radio frequency beat note on the backward and on the forward scan as function of the detuning. \textbf{c}. Power spectral density (PSD) and the reconstructed temporal waveform of the soliton state obtained on the forward wavelength scan. \textbf{d}. Interferogram (gray) and output intensity (orange) when the soliton generator is turned on and off. The same state is recovered at each power cycle following the on-off algorithm that involves adjusting the biases of RT, FP, and HT. \textbf{e}. PSD in the soliton state, acquired at equally distributed time intervals, showing an uninterrupted and intact soliton state over a span of four hours.}
     \label{fig:integrated_pump}
\end{figure*}

Pulsed operation of a semiconductor laser requires both a mechanism for pulse formation and a mechanism for pulse stabilization. In SMMLs the pulse is formed and stabilized by an intracavity saturable absorber, that opens a short temporal window of net gain only for the optical pulse, and attenuates the background, preventing its growth and pulse destabilization (Fig.~\ref{fig:pulse_generators}\textbf{b}). 
A competing destabilizing effect of the optical gain, that acts such as to amplify the noise spikes on the low-intensity background, is weak if the active medium is slow as the gain does not have the time to recover after being depleted by the pulse. In contrast, in lasers with a fast gain recovery a saturable absorber is not able to stabilize the pulse. It is for this reason, that mid-infrared QCLs, where the gain recovery time is typically less than a picosecond, up to now could not be passively mode-locked to emit pulses. In free-running standing-wave cavity Fabry-Perot QCLs, the locked frequency comb state corresponds to a frequency-modulated (FM) wave with a nearly flat intensity in the time domain~\cite{Singleton2018EvidenceLasers,Opacak2019TheoryNonlinearity,Burghoff2020UnravelingTheory,Taschler2023AsynchronousCombs}. In traveling-wave ring QCLs frequency comb operation yields gray or dark --- instead of bright --- pulses, on a high intensity background~\cite{Meng2021DissipativeLasers, Opacak2023Nozaki-BekkiSolitons}. 

An alternative pulse stabilizing mechanism is optical bistability that may occur in resonators where the optical susceptibility of the host material depends on the intracavity field intensity. In passive resonators, the susceptibility can be taken as purely real in the limit of large detuning of the frequency of optical field from the material absorption frequency. The intensity-dependent real part of susceptibility, the refractive index, is at the origin of stable soliton formation in passive systems with no optical gain, such as fiber loop resonators~\cite{Leo2010TemporalBuffer} and waveguide-based microresonators~\cite{Kippenberg2018DissipativeMicroresonators}. In these systems, a coherent CW laser field injected into the resonator may break up into one or multiple pulses that are formed by the competing action of the nonlinearity and the cavity dispersion. The typical soliton excitation scheme involves scanning the wavelength of the external pump laser through resonance from the blue-detuned side to the red-detuned side (forward scan)~\cite{Herr2013TemporalMicroresonators}. During the scan, the CW pump undergoes an instability and forms one or several cavity solitons (Fig.~\ref{fig:pulse_generators}\textbf{c}). Once excited, these solitons remain stable against noise when the driving field is effectively red-detuned from the cavity resonance. 
Solitons cannot be excited by scanning the laser wavelength from the red-detuned to the blue-detuned side of the resonance (backward scan) --- a manifestation of the bistable behavior of this optical system (Fig.~\ref{fig:pulse_generators}\textbf{c}). 

Here we experimentally demonstrate hybrid nonlinear photonic devices by applying the framework of soliton generation in passive Kerr microresonators to a semiconductor laser above its threshold. We consider a traveling-wave ring cavity filled with the optical gain medium where the bistability stems from the resonant third-order nonlinearity. In a laser cavity the medium susceptibility is as well intensity-dependent, by virtue of gain saturation. Temporal changes in the field intensity lead to changes in the imaginary part of the susceptibility. The real part of the susceptibility, inherently dependent on its imaginary part, will equally follow changes in the intracavity field intensity. This behaviour leads to the refractive nonlinearity that is akin to the Kerr nonlinearity of passive microresonators. In a semiconductor with fast gain, the refractive nonlinearity can be quantified by the laser linewidth enhancement factor (LEF)~\cite{Henry1982TheoryLasers}, which in QCLs has been shown to originate from the Bloch gain~\cite{Opacak2021FrequencyNonlinearity}. Unlike the bulk crystal Kerr nonlinearity, the nonlinearity induced by the LEF is strongly dispersive, and the nonlinear coefficient may even attain both positive and negative values at different frequencies within the gain bandwidth~\cite{Opacak2021SpectrallyComb}. 

The nonlinear active resonators have two important differences with respect to passive Kerr resonators. First, whereas the gain medium is fast, with the typical gain recovery time on the order of a few hundred femtoseconds, the nonlinearity itself that arises from the saturation of the fast gain is still slower than the almost instantaneous Kerr nonlinearity of passive resonators. The slow nonlinearity limits the bandwidth over which it acts, albeit stronger than a fast nonlinearity~\cite{Khurgin2023NonlinearTime}. Second, unlike in an optically pumped microresonator, the optical pump field is generated directly inside the ring laser cavity above its threshold. In a free-running laser, the pump frequency is tightly linked to the position of the laser cold cavity resonance, and its detuning from the resonance cannot be freely controlled; the laser operates at an optimal wavelength located between the peak of the cavity resonance and the peak of the saturated gain~\cite{Shimoda1986IntroductionPhysics}. As a result, the bistability cannot be reached without a control over the field detuning, and thus, bright solitons cannot be generated in free-running lasers~\cite{Opacak2023Nozaki-BekkiSolitons}. 

Injecting an external control optical field from a wavelength-tunable laser enables the bistability and allows one to overcome this limitation. In such an externally driven scheme, the intracavity field will injection lock to the drive field when the frequency detuning between the two is sufficiently small (Fig.~\ref{fig:pulse_generators}\textbf{d}). The intracavity field frequency will follow the frequency of the drive field --- as long as the drive field remains within the finite locking range of frequencies --- thus effectively decoupling it from the cavity resonance. Previously we theoretically proposed that such a scheme would allow the generation of optical solitons in a laser once the intracavity field is injection locked to the drive field and the intensity of the driving field is first increased, and then decreased~\cite{Columbo2021UnifyingLasers}. Alternatively, akin to the soliton generation scheme of Kerr microresonators, we find that solitons can be generated by performing wavelength scans of the drive field through the lasing resonance (Fig.~\ref{fig:pulse_generators}\textbf{d}). The intracavity field injection locks to the drive and generates one or several cavity solitons --- in direct analogy with passive Kerr microresonators.


We generate solitons in active racetrack (RT) QC resonators, operated above their lasing threshold~\cite{Kazakov2022SemiconductorCouplers}. The RT operates as a single-mode laser when a drive signal injected through an on-chip directional coupler from an external cavity laser (ECL) is detuned far off the RT lasing resonance (Fig.~\ref{fig:forward_backward_exp}\textbf{a}). The RT QCL operates in a unidirectional regime --- an essential prerequisite to the soliton generation in active laser resonators~\cite{Columbo2021UnifyingLasers}. Furthermore, unidirectional operation makes the ring laser essentially feedback-insensitive --- lifting the requirement of an optical isolator. The back-reflected wave quickly dies out inside of the resonator since the gain is strongly saturated by the forward propagating intracavity field~\cite{Opacak2023Nozaki-BekkiSolitons}. By sweeping the current of the integrated heater (HT) we can move the RT lasing resonance with respect to the drive to perform the wavelength scans~\cite{Joshi2016ThermallyMicroresonators}. In the backward wavelength scan (RT lasing resonance scanning from the red-detuned side to the blue-detuned side of the ECL frequency), the RT remains in single-mode operation, when being both in resonance and out of resonance (Fig.~\ref{fig:forward_backward_exp}\textbf{b}, \textbf{c}). On the forward wavelength scan (RT tuned from blue-detuned to red-detuned), the output intensity features a characteristic step-like drop, at which a soliton is generated (Fig.~\ref{fig:forward_backward_exp}\textbf{d}, \textbf{g}). This step is a defining feature of soliton generation in both passive Kerr resonators and fiber resonators~\cite{Herr2013TemporalMicroresonators,Englebert2021TemporalResonator}. Importantly, the RT remains injection-locked to the ECL drive field throughout the backward scan, (Fig.~\ref{fig:forward_backward_exp}\textbf{e}). The soliton spectrum is comprised of a strong pump and drive field line and a family of equidistant modes on one side of the pump line (Fig.~\ref{fig:forward_backward_exp}\textbf{h}). Performing identical backward scans multiple times reveals the existence of several possible steps in the output intensity (Fig.~\ref{fig:forward_backward_exp}\textbf{f}) signifying the multistability of the soliton states --- a feature inherent to solitons. Single- and multisoliton states can be excited in a stochastic manner by performing the same scan multiple times (Fig.~\ref{fig:forward_backward_exp}\textbf{h}, \textbf{i}). 

We found the salient spectral asymmetry to be a general feature of our active resonator solitons. It stands in a strong contrast with the highly symmetric spectra of passive resonator solitons. While the spectral asymmetry is not captured by the GLLE model, we hint towards three factors that may be causing it. First, taking into account higher order dispersion in the GLLE yields asymmetric states in simulations. Second, the highly dispersive nature of the nonlinearity induced by the LEF may as well a contributing factor to the observed spectral shape. Third, the GLLE assumes laser operation with no detuning of the pump field from the gain peak. In reality, it is hardly ever the case, and the intracavity field is likely detuned from the gain peak by a finite value. We have found the detuning of the drive laser frequency from the gain peak to be a contributing factor to the spectral asymmetry. 


One impeding characteristic of the driven cavity solitons is the presence of a strong background driving field that manifests itself in a mode $20-30$ dB greater than the rest of the soliton spectrum. This feature necessitates the use of detectors with high dynamic range and may cause gain depletion of optical amplifiers by the unnecessarily strong laser line. 
In passive Kerr resonators the pump is typically filtered out by an external fiber Bragg grating~\cite{Trocha2018UltrafastCombs} or by outcoupling the intracavity field through an add-drop waveguide~\cite{Wang2016IntracavityRegime}. The first approach is not compatible with photonic integration as the mid-infrared optics is based on free-space components. The second approach would not work with the active resonators presented here, since the pump field that drives soliton formation, despite being controlled by an externally injected field, arises directly inside the cavity and would be equally outcoupled through the drop port. 

Instead, we achieve pump suppression with a photonic integrated notch filter. An active ring resonator driven below the lasing threshold can act as such a filter, where the coupling condition, the quality factor, and the resonance frequency can be tuned independently via electrical pumping~\cite{Kazakov2022SemiconductorCouplers}. Such a notch filter targeting only one laser line can be naturally integrated on the same chip with the soliton generator. In our implementation, both the soliton generator and the notch filter are identical racetrack resonators connected via a bus waveguide. The resonators have nominally equal free spectral range (FSR) which can be finely tuned by adjusting the pump current in both resonators and the integrated heaters (Fig.~\ref{fig:integrated_filter}\textbf{a}). The soliton is generated by an off-chip ECL in the first RT QCL driven above threshold, as in Fig.~\ref{fig:forward_backward_exp}, and the pump line is subsequently filtered by the second RT QCL, operated below threshold (Fig.~\ref{fig:integrated_filter}\textbf{a}). Fig.~\ref{fig:integrated_filter}\textbf{b} shows the spectrogram of the output of this chip while tuning the bias of the notch filter. The filter is close to the critical coupling, and its FSR is slightly detuned from the intermode spacing of the soliton frequency comb, so that, relying on a Vernier effect, it filters out one comb line at each current setting. The filter suppresses the pump line by 45 dB when tuned in resonance with the driving field (Fig.~\ref{fig:integrated_filter}\textbf{c}). Fig.~\ref{fig:integrated_filter}\textbf{d} shows the spectra of the output state when the filter is turned off and on, signifying the highly selective suppression of the driving field, while mildly affecting the rest of the comb spectrum.  A soliton on a high-intensity background turns into a background-free bright pulse when the filter is switched on.  (Fig.~\ref{fig:integrated_filter}\textbf{e}).



The requirement of an additional external laser to drive the formation of solitons and the reliance on the precise optical alignment of the system would hinder the practical deployment of these devices in real applications. To address this issue we further demonstrate a mid-infrared photonic chip where the pump laser is monolithically integrated with the active racetrack resonator. Fig.~\ref{fig:integrated_pump}\textbf{a} shows such a chip where the drive signal is derived from a Fabry-Perot (FP) cavity laser that is butt-coupled to the waveguide directional coupler. Bringing the FP and the RT above their lasing thresholds, both in single-mode operation, and tuning the current of the heater (HT) integrated next to RT, we perform the wavelength sweep of the FP through the RT resonance in the same way as was done with an external drive laser. On a red-detuned to blue-detuned backward scan over a finite range of detuning the output intensity exhibits low-frequency fluctuations (Fig.~\ref{fig:integrated_pump}\textbf{b}), strongly reminiscent of the modulation instability comb regime in passive Kerr resonators~\cite{Herr2013TemporalMicroresonators}. As in Kerr resonators, the associated RF spectrum of the intermode beat note features multiple frequency tones. On the forward scan (blue-detuned to red-detuned), soliton formation is associated with the disappearance of the intensity fluctuations in the laser output, while the driver laser is still in resonance, and the emergence of a stable microwave tone at the roundtrip frequency, signifying the high degree of the phase coherence of the multimode state (Fig.~\ref{fig:integrated_pump}\textbf{b}). The spectrum of the soliton state features tens of lines, all mutually locked, giving rise to a stable bright pulse emission on top of the CW background (Fig.~\ref{fig:integrated_pump}\textbf{c}). 

Another superior attribute of the integrated soliton generator, besides its compactness, is the long-term stability of the soliton state and its ability to recover upon power cycling. We demonstrate turnkey soliton generation and perfect state recovery in a sequence of power cycles by switching on and off the drive currents of the integrated components in a prescribed manner (Fig.~\ref{fig:integrated_pump}\textbf{d}). The soliton state initiated in this way remains stable for hours of continuous operation (Fig.~\ref{fig:integrated_pump}\textbf{e}). These features of the fully integrated system are as well prevalent in Si$_3$N$_4$ Kerr resonators heterogeneously integrated with III-V semiconductor pump lasers~\cite{Xiang2021LaserSilicon}. 



The demonstrated devices are prototypical components for more elaborate mid-infrared photonic integrated architectures for short pulse generation. We showed the first DC-driven, chip-scale pulse generator in this wavelength range. A straightforward, at the same time, key improvement will be to raise the peak pulse intensity to the level suitable for supercontinuum generation in nonlinear crystals, chalcogenide glass fibers, and passive III-V/Si/Ge waveguides~\cite{Bres2023SupercontinuumPerspectives}. Power gains can be achieved by improved chip thermal management (buried heterostructure regrowth, epidown mounting) and by combining the designs shown in Fig.~\ref{fig:integrated_filter} and Fig.~\ref{fig:integrated_pump} with an additional integration of a semiconductor optical amplifier at the end of the waveguide coupler. Although the current demonstration is performed at 8.3 $\upmu$m, our approach generalizes throughout the mid-infrared ($4-12$ $\upmu$m) by engineering the QCL active region, and potentially, to shorter wavelengths using other semiconductor laser gain media --- interband cascade, quantum well, quantum dot, and quantum dash. Soliton mode-locking reliant on a bistability may be a compelling alternative to passive mode-locking with a saturable absorber in these laser systems as the absorber induces additional cavity loss that limits the average optical power and the pulse energy. 

The maturity of both semiconductor lasers, active resonators, and nonlinear Kerr cavities, passive resonators, invites one to consider the shift in the near future towards hybrid passive-active nonlinear integrated systems. On one hand, passive Si$_3$N$_4$ nonlinear resonators can be endowed with active functions --- amplification and lasing --- by ion implantation~\cite{Liu2022AAmplifier, Gaafar2023FemtosecondChip} or by heterogeneous and monolithic integration of the III-V active regions and passive waveguides~\cite{Xiang20233DPhotonics}. On the other hand, as we show here, concepts initially coined in the context of passive cavities can be extended to resonators with gain, allowing observation of new phenomena, such as direct on-chip bright pulse generation, up to now regarded as improbable in QCLs due to the absence of a suitable mode-locking mechanism. 

In this work, all components, including the routing waveguides, are made of active media, necessitating electrical biasing to reduce optical loss. Further monolithic integration of III-V passive semiconductor waveguides with QC active media~\cite{Jung2019HomogeneousPlatform,Wang2022MonolithicWaveguides} should enable larger-scale integrated architectures for linear and nonlinear absorption spectroscopy, absolute time and frequency metrology in the $4-12$ $\upmu$m window, and free-space optical communications~\cite{Dely202210Optoelectronics}. For instance, on-chip dual-comb QCL spectrometers is a long-standing vision proponed multiple times the past~\cite{Villares2015On-chipCombs}. The fact that it is now possible to have multiple active components on a chip --- including lasers above threshold --- biased independently and working simultaneously, each fulfilling their own function, is another non-trivial and important result of this work. The intrinsic immunity of the ring resonator based QCL frequency combs to delayed optical feedback~\cite{Opacak2023Nozaki-BekkiSolitons} further justifies the on-chip integration of such components to create functional spectroscopic and metrological systems, eliminating the need for an optical isolator.  

\section*{Methods}

\textbf{Generalized model}. Numerical simulations are carried out by integrating the GLLE model on a regular CPU. The simulation parameters are identical to those in Ref.~\cite{Prati2021SolitonSignal}. The simulations of both passive and active devices based on the GLLE, were performed using two different integration schemes, verifying the robustness of our theoretical predictions. One code is based on the Exponential Time Stepping scheme~\cite{Cox2002ExponentialSystems} with  large spatial grids (500-10000 points) and time steps optimized for stability, convergence and numerical efficiency. The second scheme is based on the modal decomposition of the electric field $E$ in terms of the longitudinal modes of the cold cavity. The field is expanded into $2N+1$ modes (the central one and $N$ modes for each side of the spectrum). Typically, $N=100$. 

We further provide an application (MAC OS and Windows) for a custom GUI-enhanced GLLE solver that we used for rapid prototyping of the longer parameter sweeps.

\textbf{Device fabrication and operation}. The lasers emit at around 8.3 $\upmu$m and have a structure consisting of GaInAs/AlInAs layers on an InP substrate. The active region band structure design is based on a single-phonon continuum depopulation scheme. The waveguides are dry etched (6 $\upmu$m depth) using the standard ridge process with optical lithography. The dry etch results in nearly 90$^{\circ}$ vertical waveguide sidewalls. Waveguide width is 10 $\upmu$m, the curved section of the racetrack is a semicircle with a radius of 500 $\upmu$m, and the length of the straight section is $785$ $\upmu$m for the devices in Fig.~\ref{fig:forward_backward_exp} and \ref{fig:integrated_filter} (FSR 18.6 GHz), and $1.5$ mm for the device in Fig.~\ref{fig:integrated_pump} (FSR 13.8 GHz). The air gap in the directional coupler section is 1 $\upmu$m wide. The facets of the Fabry-Perot pump lasers and of the waveguide couplers are left as cleaved or are dry etched in places where the optical coupling occurs on the chip. Laser dies are indium soldered episide up on copper carriers and individual contact sections are wirebonded to PCBs. All chips are placed on temperature-stabilized heat sinks at 16$^{\circ}$C. All components (laser resonators, waveguide couplers, heaters) are individually biased with low-noise current sources (Wavelength Electronics QCL1500 or QCL2000).  

\textbf{Optical injection setup}. The coherent optical drive signal is derived from a custom-built external cavity tunable QCL (ECL) by DRS Daylight Solutions. The beam from the ECL is focused onto the WG facet with an aspheric antireflection coated lens (NA$=0.56$) and the output radiation is collected at the opposite WG facet with an identical lens. Part of the output is focused directly onto an MCT photodetector (VIGO photonics, 300 MHz cutoff) for the output power/transmission measurement. Another part is sent through a Michelson interferometer onto an MCT photodetector (VIGO photonics, 1 GHz cutoff) for acquisition of an interferogram and of an optical spectrum. The detector outputs are sampled at 1.8 GSa/s (output intensity) and 7 MSa/s (interferogram) using data acquisition (DAQ) and oscilloscope (OSC) modules of the Zurich Instruments MF and UHF lock-in amplifiers. We control the resonance detuning and perform the resonance sweep scans by fixig the wavelength of the ECL and applying a triangular wave modulation with a function generator (SWEEP box) to the electrical drive current of heater integrated next to the racetrack resonator. An RF spectrum of the intermode beatnote is acquired by bringing an RF probe in contact with a laser die. The temporal waveforms are reconstructed using SWIFTS~\cite{Burghoff2020SensitivitySpectroscopy}.

\section*{Acknowledgements}
This material is based on work supported by the National Science Foundation under Grant No. ECCS-2221715. T. P. Letsou thanks the Department of Defense (DoD) through the National Defense Science and Engineering Graduate (NDSEG) Fellowship Program. N. Opa$\mathrm{\check{c}}$ak and B. Schwarz are supported by the European Research Council (853014). D. Kazakov and T. Letsou thank Prof. Kiyoul Yang and Tony Song (Harvard University) for enlightening discussions. 

\section*{Data availability}
The data that support the findings of this study are available upon request from the authors.

\section*{Author contributions}
M. Piccardo and D.K. conceived the project. D.K. designed the devices, M.B., S.D.C., B.S., fabricated them. D.K. designed the experiments. D.K., T.P.L., N.O, and P.R. carried out measurements and analyzed data. L.L.C, M.B., F.P., and L.A.L. developed the theory of driven active resonators and carried out numerical simulations. D.K. and M. Piccardo developed the interactive GLLE solver. M. Pushkarsky, D.C., and T.D. manufactured an external cavity tunable laser. All authors participated in the production of the manuscript. B.S., M. Piccardo, and F.C. supervised the project.   

\section*{Competing interests}
The authors declare the existence of a financial competing interest. The Office of Technology Development of Harvard University has begun the process of filing a patent application based on the materials of this work and is exploring commercialization opportunities for the presented technology.




\bibliography{references}

\begin{thebibliography}{52}%
\makeatletter
\providecommand \@ifxundefined [1]{%
 \@ifx{#1\undefined}
}%
\providecommand \@ifnum [1]{%
 \ifnum #1\expandafter \@firstoftwo
 \else \expandafter \@secondoftwo
 \fi
}%
\providecommand \@ifx [1]{%
 \ifx #1\expandafter \@firstoftwo
 \else \expandafter \@secondoftwo
 \fi
}%
\providecommand \natexlab [1]{#1}%
\providecommand \enquote  [1]{``#1''}%
\providecommand \bibnamefont  [1]{#1}%
\providecommand \bibfnamefont [1]{#1}%
\providecommand \citenamefont [1]{#1}%
\providecommand \href@noop [0]{\@secondoftwo}%
\providecommand \href [0]{\begingroup \@sanitize@url \@href}%
\providecommand \@href[1]{\@@startlink{#1}\@@href}%
\providecommand \@@href[1]{\endgroup#1\@@endlink}%
\providecommand \@sanitize@url [0]{\catcode `\\12\catcode `\$12\catcode `\&12\catcode `\#12\catcode `\^12\catcode `\_12\catcode `\%12\relax}%
\providecommand \@@startlink[1]{}%
\providecommand \@@endlink[0]{}%
\providecommand \url  [0]{\begingroup\@sanitize@url \@url }%
\providecommand \@url [1]{\endgroup\@href {#1}{\urlprefix }}%
\providecommand \urlprefix  [0]{URL }%
\providecommand \Eprint [0]{\href }%
\providecommand \doibase [0]{https://doi.org/}%
\providecommand \selectlanguage [0]{\@gobble}%
\providecommand \bibinfo  [0]{\@secondoftwo}%
\providecommand \bibfield  [0]{\@secondoftwo}%
\providecommand \translation [1]{[#1]}%
\providecommand \BibitemOpen [0]{}%
\providecommand \bibitemStop [0]{}%
\providecommand \bibitemNoStop [0]{.\EOS\space}%
\providecommand \EOS [0]{\spacefactor3000\relax}%
\providecommand \BibitemShut  [1]{\csname bibitem#1\endcsname}%
\let\auto@bib@innerbib\@empty
\bibitem [{\citenamefont {Chang}\ \emph {et~al.}(2022)\citenamefont {Chang}, \citenamefont {Liu},\ and\ \citenamefont {Bowers}}]{Chang2022IntegratedTechnologies}%
  \BibitemOpen
  \bibfield  {author} {\bibinfo {author} {\bibfnamefont {L.}~\bibnamefont {Chang}}, \bibinfo {author} {\bibfnamefont {S.}~\bibnamefont {Liu}},\ and\ \bibinfo {author} {\bibfnamefont {J.~E.}\ \bibnamefont {Bowers}},\ }\bibfield  {title} {\bibinfo {title} {{Integrated optical frequency comb technologies}},\ }\href {https://doi.org/10.1038/s41566-021-00945-1} {\bibfield  {journal} {\bibinfo  {journal} {Nature Photonics}\ }\textbf {\bibinfo {volume} {16}},\ \bibinfo {pages} {95} (\bibinfo {year} {2022})}\BibitemShut {NoStop}%
\bibitem [{\citenamefont {Kippenberg}\ \emph {et~al.}(2018)\citenamefont {Kippenberg}, \citenamefont {Gaeta}, \citenamefont {Lipson},\ and\ \citenamefont {Gorodetsky}}]{Kippenberg2018DissipativeMicroresonators}%
  \BibitemOpen
  \bibfield  {author} {\bibinfo {author} {\bibfnamefont {T.~J.}\ \bibnamefont {Kippenberg}}, \bibinfo {author} {\bibfnamefont {A.~L.}\ \bibnamefont {Gaeta}}, \bibinfo {author} {\bibfnamefont {M.}~\bibnamefont {Lipson}},\ and\ \bibinfo {author} {\bibfnamefont {M.~L.}\ \bibnamefont {Gorodetsky}},\ }\bibfield  {title} {\bibinfo {title} {{Dissipative Kerr solitons in optical microresonators}},\ }\bibfield  {journal} {\bibinfo  {journal} {Science}\ }\textbf {\bibinfo {volume} {361}},\ \href {https://doi.org/10.1126/science.aan8083} {10.1126/science.aan8083} (\bibinfo {year} {2018})\BibitemShut {NoStop}%
\bibitem [{\citenamefont {Liu}\ \emph {et~al.}(2019)\citenamefont {Liu}, \citenamefont {Wu}, \citenamefont {Jung}, \citenamefont {Norman}, \citenamefont {Kennedy}, \citenamefont {Tsang}, \citenamefont {Gossard},\ and\ \citenamefont {Bowers}}]{Liu2019High-channel-countCapacity}%
  \BibitemOpen
  \bibfield  {author} {\bibinfo {author} {\bibfnamefont {S.}~\bibnamefont {Liu}}, \bibinfo {author} {\bibfnamefont {X.}~\bibnamefont {Wu}}, \bibinfo {author} {\bibfnamefont {D.}~\bibnamefont {Jung}}, \bibinfo {author} {\bibfnamefont {J.~C.}\ \bibnamefont {Norman}}, \bibinfo {author} {\bibfnamefont {M.~J.}\ \bibnamefont {Kennedy}}, \bibinfo {author} {\bibfnamefont {H.~K.}\ \bibnamefont {Tsang}}, \bibinfo {author} {\bibfnamefont {A.~C.}\ \bibnamefont {Gossard}},\ and\ \bibinfo {author} {\bibfnamefont {J.~E.}\ \bibnamefont {Bowers}},\ }\bibfield  {title} {\bibinfo {title} {{High-channel-count 20 GHz passively mode-locked quantum dot laser directly grown on Si with 4.1 Tbit/s transmission capacity}},\ }\href {https://doi.org/10.1364/OPTICA.6.000128} {\bibfield  {journal} {\bibinfo  {journal} {Optica}\ }\textbf {\bibinfo {volume} {6}},\ \bibinfo {pages} {128} (\bibinfo {year} {2019})}\BibitemShut {NoStop}%
\bibitem [{\citenamefont {Hillbrand}\ \emph {et~al.}(2019)\citenamefont {Hillbrand}, \citenamefont {Beiser}, \citenamefont {Andrews}, \citenamefont {Detz}, \citenamefont {Schade}, \citenamefont {Weih}, \citenamefont {H{\"{o}}fling}, \citenamefont {Schwarz},\ and\ \citenamefont {Strasser}}]{Hillbrand2019PicosecondLaser}%
  \BibitemOpen
  \bibfield  {author} {\bibinfo {author} {\bibfnamefont {J.}~\bibnamefont {Hillbrand}}, \bibinfo {author} {\bibfnamefont {M.}~\bibnamefont {Beiser}}, \bibinfo {author} {\bibfnamefont {A.~M.}\ \bibnamefont {Andrews}}, \bibinfo {author} {\bibfnamefont {H.}~\bibnamefont {Detz}}, \bibinfo {author} {\bibfnamefont {A.}~\bibnamefont {Schade}}, \bibinfo {author} {\bibfnamefont {R.}~\bibnamefont {Weih}}, \bibinfo {author} {\bibfnamefont {S.}~\bibnamefont {H{\"{o}}fling}}, \bibinfo {author} {\bibfnamefont {B.}~\bibnamefont {Schwarz}},\ and\ \bibinfo {author} {\bibfnamefont {G.}~\bibnamefont {Strasser}},\ }\bibfield  {title} {\bibinfo {title} {{Picosecond pulses from a mid-infrared interband cascade laser}},\ }\href {https://doi.org/10.1364/OPTICA.6.001334} {\bibfield  {journal} {\bibinfo  {journal} {Optica}\ }\textbf {\bibinfo {volume} {6}},\ \bibinfo {pages} {1334} (\bibinfo {year} {2019})}\BibitemShut {NoStop}%
\bibitem [{\citenamefont {Ren}\ \emph {et~al.}(2023)\citenamefont {Ren}, \citenamefont {Dong},\ and\ \citenamefont {Burghoff}}]{Ren2023IntegratedRoadmap}%
  \BibitemOpen
  \bibfield  {author} {\bibinfo {author} {\bibfnamefont {D.}~\bibnamefont {Ren}}, \bibinfo {author} {\bibfnamefont {C.}~\bibnamefont {Dong}},\ and\ \bibinfo {author} {\bibfnamefont {D.}~\bibnamefont {Burghoff}},\ }\bibfield  {title} {\bibinfo {title} {{Integrated nonlinear photonics in the longwave-infrared: A roadmap}},\ }\href {https://doi.org/10.1557/S43579-023-00435-1} {\bibfield  {journal} {\bibinfo  {journal} {MRS Communications}\ ,\ \bibinfo {pages} {1}} (\bibinfo {year} {2023})}\BibitemShut {NoStop}%
\bibitem [{\citenamefont {Marin-Palomo}\ \emph {et~al.}(2019)\citenamefont {Marin-Palomo}, \citenamefont {Kemal}, \citenamefont {Trocha}, \citenamefont {Wolf}, \citenamefont {Merghem}, \citenamefont {Lelarge}, \citenamefont {Ramdane}, \citenamefont {Freude}, \citenamefont {Randel},\ and\ \citenamefont {Koos}}]{Marin-Palomo2019Comb-basedDiode}%
  \BibitemOpen
  \bibfield  {author} {\bibinfo {author} {\bibfnamefont {P.}~\bibnamefont {Marin-Palomo}}, \bibinfo {author} {\bibfnamefont {J.~N.}\ \bibnamefont {Kemal}}, \bibinfo {author} {\bibfnamefont {P.}~\bibnamefont {Trocha}}, \bibinfo {author} {\bibfnamefont {S.}~\bibnamefont {Wolf}}, \bibinfo {author} {\bibfnamefont {K.}~\bibnamefont {Merghem}}, \bibinfo {author} {\bibfnamefont {F.}~\bibnamefont {Lelarge}}, \bibinfo {author} {\bibfnamefont {A.}~\bibnamefont {Ramdane}}, \bibinfo {author} {\bibfnamefont {W.}~\bibnamefont {Freude}}, \bibinfo {author} {\bibfnamefont {S.}~\bibnamefont {Randel}},\ and\ \bibinfo {author} {\bibfnamefont {C.}~\bibnamefont {Koos}},\ }\bibfield  {title} {\bibinfo {title} {{Comb-based WDM transmission at 10 Tbit/s using a DC-driven quantum-dash mode-locked laser diode}},\ }\href {https://doi.org/10.1364/OE.27.031110} {\bibfield  {journal} {\bibinfo  {journal} {Optics Express}\ }\textbf {\bibinfo {volume} {27}},\ \bibinfo {pages} {31110} (\bibinfo {year} {2019})}\BibitemShut {NoStop}%
\bibitem [{\citenamefont {Dausinger}\ \emph {et~al.}(2004)\citenamefont {Dausinger}, \citenamefont {Lubatschowski},\ and\ \citenamefont {Lichtner}}]{Dausinger2004FemtosecondApplications}%
  \BibitemOpen
  \bibfield  {author} {\bibinfo {author} {\bibfnamefont {F.}~\bibnamefont {Dausinger}}, \bibinfo {author} {\bibfnamefont {H.}~\bibnamefont {Lubatschowski}},\ and\ \bibinfo {author} {\bibfnamefont {F.}~\bibnamefont {Lichtner}},\ }\href {https://doi.org/10.1007/B96440} {\emph {\bibinfo {title} {{Femtosecond Technology for Technical and Medical Applications}}}},\ edited by\ \bibinfo {editor} {\bibfnamefont {F.}~\bibnamefont {Dausinger}}, \bibinfo {editor} {\bibfnamefont {H.}~\bibnamefont {Lubatschowski}},\ and\ \bibinfo {editor} {\bibfnamefont {F.}~\bibnamefont {Lichtner}},\ Topics in Applied Physics\ (\bibinfo  {publisher} {Springer Berlin Heidelberg},\ \bibinfo {address} {Berlin, Heidelberg},\ \bibinfo {year} {2004})\BibitemShut {NoStop}%
\bibitem [{\citenamefont {Riemensberger}\ \emph {et~al.}(2020)\citenamefont {Riemensberger}, \citenamefont {Lukashchuk}, \citenamefont {Karpov}, \citenamefont {Weng}, \citenamefont {Lucas}, \citenamefont {Liu},\ and\ \citenamefont {Kippenberg}}]{Riemensberger2020MassivelyMicrocomb}%
  \BibitemOpen
  \bibfield  {author} {\bibinfo {author} {\bibfnamefont {J.}~\bibnamefont {Riemensberger}}, \bibinfo {author} {\bibfnamefont {A.}~\bibnamefont {Lukashchuk}}, \bibinfo {author} {\bibfnamefont {M.}~\bibnamefont {Karpov}}, \bibinfo {author} {\bibfnamefont {W.}~\bibnamefont {Weng}}, \bibinfo {author} {\bibfnamefont {E.}~\bibnamefont {Lucas}}, \bibinfo {author} {\bibfnamefont {J.}~\bibnamefont {Liu}},\ and\ \bibinfo {author} {\bibfnamefont {T.~J.}\ \bibnamefont {Kippenberg}},\ }\bibfield  {title} {\bibinfo {title} {{Massively parallel coherent laser ranging using a soliton microcomb}},\ }\href {https://www.nature.com/articles/s41586-020-2239-3} {\bibfield  {journal} {\bibinfo  {journal} {Nature}\ }\textbf {\bibinfo {volume} {581}},\ \bibinfo {pages} {164} (\bibinfo {year} {2020})}\BibitemShut {NoStop}%
\bibitem [{\citenamefont {Dudley}\ \emph {et~al.}(2006)\citenamefont {Dudley}, \citenamefont {Genty},\ and\ \citenamefont {Coen}}]{Dudley2006SupercontinuumFiber}%
  \BibitemOpen
  \bibfield  {author} {\bibinfo {author} {\bibfnamefont {J.~M.}\ \bibnamefont {Dudley}}, \bibinfo {author} {\bibfnamefont {G.}~\bibnamefont {Genty}},\ and\ \bibinfo {author} {\bibfnamefont {S.}~\bibnamefont {Coen}},\ }\bibfield  {title} {\bibinfo {title} {{Supercontinuum generation in photonic crystal fiber}},\ }\href {https://doi.org/10.1103/REVMODPHYS.78.1135/FIGURES/28/MEDIUM} {\bibfield  {journal} {\bibinfo  {journal} {Reviews of Modern Physics}\ }\textbf {\bibinfo {volume} {78}},\ \bibinfo {pages} {1135} (\bibinfo {year} {2006})}\BibitemShut {NoStop}%
\bibitem [{\citenamefont {Takamoto}\ \emph {et~al.}(2020)\citenamefont {Takamoto}, \citenamefont {Ushijima}, \citenamefont {Ohmae}, \citenamefont {Yahagi}, \citenamefont {Kokado}, \citenamefont {Shinkai},\ and\ \citenamefont {Katori}}]{Takamoto2020TestClocks}%
  \BibitemOpen
  \bibfield  {author} {\bibinfo {author} {\bibfnamefont {M.}~\bibnamefont {Takamoto}}, \bibinfo {author} {\bibfnamefont {I.}~\bibnamefont {Ushijima}}, \bibinfo {author} {\bibfnamefont {N.}~\bibnamefont {Ohmae}}, \bibinfo {author} {\bibfnamefont {T.}~\bibnamefont {Yahagi}}, \bibinfo {author} {\bibfnamefont {K.}~\bibnamefont {Kokado}}, \bibinfo {author} {\bibfnamefont {H.}~\bibnamefont {Shinkai}},\ and\ \bibinfo {author} {\bibfnamefont {H.}~\bibnamefont {Katori}},\ }\bibfield  {title} {\bibinfo {title} {{Test of general relativity by a pair of transportable optical lattice clocks}},\ }\href {https://doi.org/10.1038/s41566-020-0619-8} {\bibfield  {journal} {\bibinfo  {journal} {Nature Photonics}\ }\textbf {\bibinfo {volume} {14}},\ \bibinfo {pages} {411} (\bibinfo {year} {2020})}\BibitemShut {NoStop}%
\bibitem [{\citenamefont {Drake}\ \emph {et~al.}(2019)\citenamefont {Drake}, \citenamefont {Briles}, \citenamefont {Stone}, \citenamefont {Spencer}, \citenamefont {Carlson}, \citenamefont {Hickstein}, \citenamefont {Li}, \citenamefont {Westly}, \citenamefont {Srinivasan}, \citenamefont {Diddams},\ and\ \citenamefont {Papp}}]{Drake2019Terahertz-RateClockwork}%
  \BibitemOpen
  \bibfield  {author} {\bibinfo {author} {\bibfnamefont {T.~E.}\ \bibnamefont {Drake}}, \bibinfo {author} {\bibfnamefont {T.~C.}\ \bibnamefont {Briles}}, \bibinfo {author} {\bibfnamefont {J.~R.}\ \bibnamefont {Stone}}, \bibinfo {author} {\bibfnamefont {D.~T.}\ \bibnamefont {Spencer}}, \bibinfo {author} {\bibfnamefont {D.~R.}\ \bibnamefont {Carlson}}, \bibinfo {author} {\bibfnamefont {D.~D.}\ \bibnamefont {Hickstein}}, \bibinfo {author} {\bibfnamefont {Q.}~\bibnamefont {Li}}, \bibinfo {author} {\bibfnamefont {D.}~\bibnamefont {Westly}}, \bibinfo {author} {\bibfnamefont {K.}~\bibnamefont {Srinivasan}}, \bibinfo {author} {\bibfnamefont {S.~A.}\ \bibnamefont {Diddams}},\ and\ \bibinfo {author} {\bibfnamefont {S.~B.}\ \bibnamefont {Papp}},\ }\bibfield  {title} {\bibinfo {title} {{Terahertz-Rate Kerr-Microresonator Optical Clockwork}},\ }\href {https://doi.org/10.1103/PHYSREVX.9.031023/FIGURES/3/MEDIUM} {\bibfield  {journal} {\bibinfo  {journal} {Physical Review X}\ }\textbf {\bibinfo {volume} {9}},\ \bibinfo
  {pages} {031023} (\bibinfo {year} {2019})}\BibitemShut {NoStop}%
\bibitem [{\citenamefont {Spencer}\ \emph {et~al.}(2018)\citenamefont {Spencer}, \citenamefont {Drake}, \citenamefont {Briles}, \citenamefont {Stone}, \citenamefont {Sinclair}, \citenamefont {Fredrick}, \citenamefont {Li}, \citenamefont {Westly}, \citenamefont {Ilic}, \citenamefont {Bluestone}, \citenamefont {Volet}, \citenamefont {Komljenovic}, \citenamefont {Chang}, \citenamefont {Lee}, \citenamefont {Oh}, \citenamefont {Suh}, \citenamefont {Yang}, \citenamefont {Pfeiffer}, \citenamefont {Kippenberg}, \citenamefont {Norberg}, \citenamefont {Theogarajan}, \citenamefont {Vahala}, \citenamefont {Newbury}, \citenamefont {Srinivasan}, \citenamefont {Bowers}, \citenamefont {Diddams},\ and\ \citenamefont {Papp}}]{Spencer2018AnPhotonics}%
  \BibitemOpen
  \bibfield  {author} {\bibinfo {author} {\bibfnamefont {D.~T.}\ \bibnamefont {Spencer}}, \bibinfo {author} {\bibfnamefont {T.}~\bibnamefont {Drake}}, \bibinfo {author} {\bibfnamefont {T.~C.}\ \bibnamefont {Briles}}, \bibinfo {author} {\bibfnamefont {J.}~\bibnamefont {Stone}}, \bibinfo {author} {\bibfnamefont {L.~C.}\ \bibnamefont {Sinclair}}, \bibinfo {author} {\bibfnamefont {C.}~\bibnamefont {Fredrick}}, \bibinfo {author} {\bibfnamefont {Q.}~\bibnamefont {Li}}, \bibinfo {author} {\bibfnamefont {D.}~\bibnamefont {Westly}}, \bibinfo {author} {\bibfnamefont {B.~R.}\ \bibnamefont {Ilic}}, \bibinfo {author} {\bibfnamefont {A.}~\bibnamefont {Bluestone}}, \bibinfo {author} {\bibfnamefont {N.}~\bibnamefont {Volet}}, \bibinfo {author} {\bibfnamefont {T.}~\bibnamefont {Komljenovic}}, \bibinfo {author} {\bibfnamefont {L.}~\bibnamefont {Chang}}, \bibinfo {author} {\bibfnamefont {S.~H.}\ \bibnamefont {Lee}}, \bibinfo {author} {\bibfnamefont {D.~Y.}\ \bibnamefont {Oh}}, \bibinfo {author} {\bibfnamefont {M.~G.}\
  \bibnamefont {Suh}}, \bibinfo {author} {\bibfnamefont {K.~Y.}\ \bibnamefont {Yang}}, \bibinfo {author} {\bibfnamefont {M.~H.}\ \bibnamefont {Pfeiffer}}, \bibinfo {author} {\bibfnamefont {T.~J.}\ \bibnamefont {Kippenberg}}, \bibinfo {author} {\bibfnamefont {E.}~\bibnamefont {Norberg}}, \bibinfo {author} {\bibfnamefont {L.}~\bibnamefont {Theogarajan}}, \bibinfo {author} {\bibfnamefont {K.}~\bibnamefont {Vahala}}, \bibinfo {author} {\bibfnamefont {N.~R.}\ \bibnamefont {Newbury}}, \bibinfo {author} {\bibfnamefont {K.}~\bibnamefont {Srinivasan}}, \bibinfo {author} {\bibfnamefont {J.~E.}\ \bibnamefont {Bowers}}, \bibinfo {author} {\bibfnamefont {S.~A.}\ \bibnamefont {Diddams}},\ and\ \bibinfo {author} {\bibfnamefont {S.~B.}\ \bibnamefont {Papp}},\ }\bibfield  {title} {\bibinfo {title} {{An optical-frequency synthesizer using integrated photonics}},\ }\href {https://doi.org/10.1038/s41586-018-0065-7} {\bibfield  {journal} {\bibinfo  {journal} {Nature}\ }\textbf {\bibinfo {volume} {557}},\ \bibinfo {pages} {81}
  (\bibinfo {year} {2018})}\BibitemShut {NoStop}%
\bibitem [{\citenamefont {Marin-Palomo}\ \emph {et~al.}(2017)\citenamefont {Marin-Palomo}, \citenamefont {Kemal}, \citenamefont {Karpov}, \citenamefont {Kordts}, \citenamefont {Pfeifle}, \citenamefont {Pfeiffer}, \citenamefont {Trocha}, \citenamefont {Wolf}, \citenamefont {Brasch}, \citenamefont {Anderson}, \citenamefont {Rosenberger}, \citenamefont {Vijayan}, \citenamefont {Freude}, \citenamefont {Kippenberg},\ and\ \citenamefont {Koos}}]{Marin-Palomo2017Microresonator-basedCommunications}%
  \BibitemOpen
  \bibfield  {author} {\bibinfo {author} {\bibfnamefont {P.}~\bibnamefont {Marin-Palomo}}, \bibinfo {author} {\bibfnamefont {J.~N.}\ \bibnamefont {Kemal}}, \bibinfo {author} {\bibfnamefont {M.}~\bibnamefont {Karpov}}, \bibinfo {author} {\bibfnamefont {A.}~\bibnamefont {Kordts}}, \bibinfo {author} {\bibfnamefont {J.}~\bibnamefont {Pfeifle}}, \bibinfo {author} {\bibfnamefont {M.~H.}\ \bibnamefont {Pfeiffer}}, \bibinfo {author} {\bibfnamefont {P.}~\bibnamefont {Trocha}}, \bibinfo {author} {\bibfnamefont {S.}~\bibnamefont {Wolf}}, \bibinfo {author} {\bibfnamefont {V.}~\bibnamefont {Brasch}}, \bibinfo {author} {\bibfnamefont {M.~H.}\ \bibnamefont {Anderson}}, \bibinfo {author} {\bibfnamefont {R.}~\bibnamefont {Rosenberger}}, \bibinfo {author} {\bibfnamefont {K.}~\bibnamefont {Vijayan}}, \bibinfo {author} {\bibfnamefont {W.}~\bibnamefont {Freude}}, \bibinfo {author} {\bibfnamefont {T.~J.}\ \bibnamefont {Kippenberg}},\ and\ \bibinfo {author} {\bibfnamefont {C.}~\bibnamefont {Koos}},\ }\bibfield  {title} {\bibinfo
  {title} {{Microresonator-based solitons for massively parallel coherent optical communications}},\ }\href {https://doi.org/10.1038/nature22387} {\bibfield  {journal} {\bibinfo  {journal} {Nature}\ }\textbf {\bibinfo {volume} {546}},\ \bibinfo {pages} {274} (\bibinfo {year} {2017})}\BibitemShut {NoStop}%
\bibitem [{\citenamefont {Cundiff}\ and\ \citenamefont {Mukamel}(2013)}]{Cundiff2013OpticalSpectroscopy}%
  \BibitemOpen
  \bibfield  {author} {\bibinfo {author} {\bibfnamefont {S.~T.}\ \bibnamefont {Cundiff}}\ and\ \bibinfo {author} {\bibfnamefont {S.}~\bibnamefont {Mukamel}},\ }\bibfield  {title} {\bibinfo {title} {{Optical multidimensional coherent spectroscopy}},\ }\href {https://doi.org/10.1063/PT.3.2047} {\bibfield  {journal} {\bibinfo  {journal} {Physics Today}\ }\textbf {\bibinfo {volume} {66}},\ \bibinfo {pages} {44} (\bibinfo {year} {2013})}\BibitemShut {NoStop}%
\bibitem [{\citenamefont {Picqu{\'{e}}}\ and\ \citenamefont {H{\"{a}}nsch}(2019)}]{Picque2019FrequencySpectroscopy}%
  \BibitemOpen
  \bibfield  {author} {\bibinfo {author} {\bibfnamefont {N.}~\bibnamefont {Picqu{\'{e}}}}\ and\ \bibinfo {author} {\bibfnamefont {T.~W.}\ \bibnamefont {H{\"{a}}nsch}},\ }\bibfield  {title} {\bibinfo {title} {{Frequency comb spectroscopy}},\ }\href {https://doi.org/10.1038/s41566-018-0347-5} {\bibfield  {journal} {\bibinfo  {journal} {Nature Photonics}\ }\textbf {\bibinfo {volume} {13}},\ \bibinfo {pages} {146} (\bibinfo {year} {2019})}\BibitemShut {NoStop}%
\bibitem [{\citenamefont {Wang}\ \emph {et~al.}(2020)\citenamefont {Wang}, \citenamefont {Slivken}, \citenamefont {Wu}, \citenamefont {Lu},\ and\ \citenamefont {Razeghi}}]{Wang2020ContinuousOperation}%
  \BibitemOpen
  \bibfield  {author} {\bibinfo {author} {\bibfnamefont {F.}~\bibnamefont {Wang}}, \bibinfo {author} {\bibfnamefont {S.}~\bibnamefont {Slivken}}, \bibinfo {author} {\bibfnamefont {D.~H.}\ \bibnamefont {Wu}}, \bibinfo {author} {\bibfnamefont {Q.~Y.}\ \bibnamefont {Lu}},\ and\ \bibinfo {author} {\bibfnamefont {M.}~\bibnamefont {Razeghi}},\ }\bibfield  {title} {\bibinfo {title} {{Continuous wave quantum cascade lasers with 5.6 W output power at room temperature and 41{\%} wall-plug efficiency in cryogenic operation}},\ }\bibfield  {journal} {\bibinfo  {journal} {AIP Advances}\ }\textbf {\bibinfo {volume} {10}},\ \href {https://doi.org/10.1063/5.0003318/1021666} {10.1063/5.0003318/1021666} (\bibinfo {year} {2020})\BibitemShut {NoStop}%
\bibitem [{\citenamefont {Schwarz}\ \emph {et~al.}(2017)\citenamefont {Schwarz}, \citenamefont {Wang}, \citenamefont {Missaggia}, \citenamefont {Mansuripur}, \citenamefont {Chevalier}, \citenamefont {Connors}, \citenamefont {McNulty}, \citenamefont {Cederberg}, \citenamefont {Strasser},\ and\ \citenamefont {Capasso}}]{Schwarz2017Watt-LevelLaser/Detector}%
  \BibitemOpen
  \bibfield  {author} {\bibinfo {author} {\bibfnamefont {B.}~\bibnamefont {Schwarz}}, \bibinfo {author} {\bibfnamefont {C.~A.}\ \bibnamefont {Wang}}, \bibinfo {author} {\bibfnamefont {L.}~\bibnamefont {Missaggia}}, \bibinfo {author} {\bibfnamefont {T.~S.}\ \bibnamefont {Mansuripur}}, \bibinfo {author} {\bibfnamefont {P.}~\bibnamefont {Chevalier}}, \bibinfo {author} {\bibfnamefont {M.~K.}\ \bibnamefont {Connors}}, \bibinfo {author} {\bibfnamefont {D.}~\bibnamefont {McNulty}}, \bibinfo {author} {\bibfnamefont {J.}~\bibnamefont {Cederberg}}, \bibinfo {author} {\bibfnamefont {G.}~\bibnamefont {Strasser}},\ and\ \bibinfo {author} {\bibfnamefont {F.}~\bibnamefont {Capasso}},\ }\bibfield  {title} {\bibinfo {title} {{Watt-Level Continuous-Wave Emission from a Bifunctional Quantum Cascade Laser/Detector}},\ }\href {https://doi.org/10.1021/ACSPHOTONICS.7B00133/ASSET/IMAGES/PH-2017-001339{\_}M002.GIF} {\bibfield  {journal} {\bibinfo  {journal} {ACS Photonics}\ }\textbf {\bibinfo {volume} {4}},\ \bibinfo {pages} {1225}
  (\bibinfo {year} {2017})}\BibitemShut {NoStop}%
\bibitem [{\citenamefont {T{\"{a}}schler}\ \emph {et~al.}(2021)\citenamefont {T{\"{a}}schler}, \citenamefont {Bertrand}, \citenamefont {Schneider}, \citenamefont {Singleton}, \citenamefont {Jouy}, \citenamefont {Kapsalidis}, \citenamefont {Beck},\ and\ \citenamefont {Faist}}]{Taschler2021FemtosecondLaser}%
  \BibitemOpen
  \bibfield  {author} {\bibinfo {author} {\bibfnamefont {P.}~\bibnamefont {T{\"{a}}schler}}, \bibinfo {author} {\bibfnamefont {M.}~\bibnamefont {Bertrand}}, \bibinfo {author} {\bibfnamefont {B.}~\bibnamefont {Schneider}}, \bibinfo {author} {\bibfnamefont {M.}~\bibnamefont {Singleton}}, \bibinfo {author} {\bibfnamefont {P.}~\bibnamefont {Jouy}}, \bibinfo {author} {\bibfnamefont {F.}~\bibnamefont {Kapsalidis}}, \bibinfo {author} {\bibfnamefont {M.}~\bibnamefont {Beck}},\ and\ \bibinfo {author} {\bibfnamefont {J.}~\bibnamefont {Faist}},\ }\bibfield  {title} {\bibinfo {title} {{Femtosecond pulses from a mid-infrared quantum cascade laser}},\ }\href {https://doi.org/10.1038/s41566-021-00894-9} {\bibfield  {journal} {\bibinfo  {journal} {Nature Photonics}\ }\textbf {\bibinfo {volume} {15}},\ \bibinfo {pages} {919} (\bibinfo {year} {2021})}\BibitemShut {NoStop}%
\bibitem [{\citenamefont {Columbo}\ \emph {et~al.}(2021)\citenamefont {Columbo}, \citenamefont {Piccardo}, \citenamefont {Prati}, \citenamefont {Lugiato}, \citenamefont {Brambilla}, \citenamefont {Gatti}, \citenamefont {Silvestri}, \citenamefont {Gioannini}, \citenamefont {Opa{\v{c}}ak}, \citenamefont {Schwarz},\ and\ \citenamefont {Capasso}}]{Columbo2021UnifyingLasers}%
  \BibitemOpen
  \bibfield  {author} {\bibinfo {author} {\bibfnamefont {L.}~\bibnamefont {Columbo}}, \bibinfo {author} {\bibfnamefont {M.}~\bibnamefont {Piccardo}}, \bibinfo {author} {\bibfnamefont {F.}~\bibnamefont {Prati}}, \bibinfo {author} {\bibfnamefont {L.~A.}\ \bibnamefont {Lugiato}}, \bibinfo {author} {\bibfnamefont {M.}~\bibnamefont {Brambilla}}, \bibinfo {author} {\bibfnamefont {A.}~\bibnamefont {Gatti}}, \bibinfo {author} {\bibfnamefont {C.}~\bibnamefont {Silvestri}}, \bibinfo {author} {\bibfnamefont {M.}~\bibnamefont {Gioannini}}, \bibinfo {author} {\bibfnamefont {N.}~\bibnamefont {Opa{\v{c}}ak}}, \bibinfo {author} {\bibfnamefont {B.}~\bibnamefont {Schwarz}},\ and\ \bibinfo {author} {\bibfnamefont {F.}~\bibnamefont {Capasso}},\ }\bibfield  {title} {\bibinfo {title} {{Unifying Frequency Combs in Active and Passive Cavities: Temporal Solitons in Externally Driven Ring Lasers}},\ }\href {https://doi.org/10.1103/PHYSREVLETT.126.173903/FIGURES/2/MEDIUM} {\bibfield  {journal} {\bibinfo  {journal} {Physical Review
  Letters}\ }\textbf {\bibinfo {volume} {126}},\ \bibinfo {pages} {173903} (\bibinfo {year} {2021})}\BibitemShut {NoStop}%
\bibitem [{\citenamefont {Lugiato}\ and\ \citenamefont {Lefever}(1987)}]{Lugiato1987SpatialSystems}%
  \BibitemOpen
  \bibfield  {author} {\bibinfo {author} {\bibfnamefont {L.~A.}\ \bibnamefont {Lugiato}}\ and\ \bibinfo {author} {\bibfnamefont {R.}~\bibnamefont {Lefever}},\ }\bibfield  {title} {\bibinfo {title} {{Spatial Dissipative Structures in Passive Optical Systems}},\ }\href {https://doi.org/10.1103/PhysRevLett.58.2209} {\bibfield  {journal} {\bibinfo  {journal} {Physical Review Letters}\ }\textbf {\bibinfo {volume} {58}},\ \bibinfo {pages} {2209} (\bibinfo {year} {1987})}\BibitemShut {NoStop}%
\bibitem [{\citenamefont {Lugiato}\ \emph {et~al.}(1988)\citenamefont {Lugiato}, \citenamefont {Oldano},\ and\ \citenamefont {Narducci}}]{Lugiato1988CooperativeLasers}%
  \BibitemOpen
  \bibfield  {author} {\bibinfo {author} {\bibfnamefont {L.~A.}\ \bibnamefont {Lugiato}}, \bibinfo {author} {\bibfnamefont {C.}~\bibnamefont {Oldano}},\ and\ \bibinfo {author} {\bibfnamefont {L.~M.}\ \bibnamefont {Narducci}},\ }\bibfield  {title} {\bibinfo {title} {{Cooperative frequency locking and stationary spatial structures in lasers}},\ }\href {https://doi.org/10.1364/JOSAB.5.000879} {\bibfield  {journal} {\bibinfo  {journal} {JOSA B}\ }\textbf {\bibinfo {volume} {5}},\ \bibinfo {pages} {879} (\bibinfo {year} {1988})}\BibitemShut {NoStop}%
\bibitem [{\citenamefont {Piccardo}\ \emph {et~al.}(2020)\citenamefont {Piccardo}, \citenamefont {Schwarz}, \citenamefont {Kazakov}, \citenamefont {Beiser}, \citenamefont {Opa{\v{c}}ak}, \citenamefont {Wang}, \citenamefont {Jha}, \citenamefont {Hillbrand}, \citenamefont {Tamagnone}, \citenamefont {Chen}, \citenamefont {Zhu}, \citenamefont {Columbo}, \citenamefont {Belyanin},\ and\ \citenamefont {Capasso}}]{Piccardo2020FrequencyTurbulence}%
  \BibitemOpen
  \bibfield  {author} {\bibinfo {author} {\bibfnamefont {M.}~\bibnamefont {Piccardo}}, \bibinfo {author} {\bibfnamefont {B.}~\bibnamefont {Schwarz}}, \bibinfo {author} {\bibfnamefont {D.}~\bibnamefont {Kazakov}}, \bibinfo {author} {\bibfnamefont {M.}~\bibnamefont {Beiser}}, \bibinfo {author} {\bibfnamefont {N.}~\bibnamefont {Opa{\v{c}}ak}}, \bibinfo {author} {\bibfnamefont {Y.}~\bibnamefont {Wang}}, \bibinfo {author} {\bibfnamefont {S.}~\bibnamefont {Jha}}, \bibinfo {author} {\bibfnamefont {J.}~\bibnamefont {Hillbrand}}, \bibinfo {author} {\bibfnamefont {M.}~\bibnamefont {Tamagnone}}, \bibinfo {author} {\bibfnamefont {W.~T.}\ \bibnamefont {Chen}}, \bibinfo {author} {\bibfnamefont {A.~Y.}\ \bibnamefont {Zhu}}, \bibinfo {author} {\bibfnamefont {L.~L.}\ \bibnamefont {Columbo}}, \bibinfo {author} {\bibfnamefont {A.}~\bibnamefont {Belyanin}},\ and\ \bibinfo {author} {\bibfnamefont {F.}~\bibnamefont {Capasso}},\ }\bibfield  {title} {\bibinfo {title} {{Frequency combs induced by phase turbulence}},\ }\href
  {https://www.nature.com/articles/s41586-020-2386-6} {\bibfield  {journal} {\bibinfo  {journal} {Nature}\ }\textbf {\bibinfo {volume} {582}},\ \bibinfo {pages} {360} (\bibinfo {year} {2020})}\BibitemShut {NoStop}%
\bibitem [{\citenamefont {Xiang}\ \emph {et~al.}(2023)\citenamefont {Xiang}, \citenamefont {Jin}, \citenamefont {Terra}, \citenamefont {Dong}, \citenamefont {Wang}, \citenamefont {Wu}, \citenamefont {Guo}, \citenamefont {Morin}, \citenamefont {Hughes}, \citenamefont {Peters}, \citenamefont {Ji}, \citenamefont {Feshali}, \citenamefont {Paniccia}, \citenamefont {Vahala},\ and\ \citenamefont {Bowers}}]{Xiang20233DPhotonics}%
  \BibitemOpen
  \bibfield  {author} {\bibinfo {author} {\bibfnamefont {C.}~\bibnamefont {Xiang}}, \bibinfo {author} {\bibfnamefont {W.}~\bibnamefont {Jin}}, \bibinfo {author} {\bibfnamefont {O.}~\bibnamefont {Terra}}, \bibinfo {author} {\bibfnamefont {B.}~\bibnamefont {Dong}}, \bibinfo {author} {\bibfnamefont {H.}~\bibnamefont {Wang}}, \bibinfo {author} {\bibfnamefont {L.}~\bibnamefont {Wu}}, \bibinfo {author} {\bibfnamefont {J.}~\bibnamefont {Guo}}, \bibinfo {author} {\bibfnamefont {T.~J.}\ \bibnamefont {Morin}}, \bibinfo {author} {\bibfnamefont {E.}~\bibnamefont {Hughes}}, \bibinfo {author} {\bibfnamefont {J.}~\bibnamefont {Peters}}, \bibinfo {author} {\bibfnamefont {Q.~X.}\ \bibnamefont {Ji}}, \bibinfo {author} {\bibfnamefont {A.}~\bibnamefont {Feshali}}, \bibinfo {author} {\bibfnamefont {M.}~\bibnamefont {Paniccia}}, \bibinfo {author} {\bibfnamefont {K.~J.}\ \bibnamefont {Vahala}},\ and\ \bibinfo {author} {\bibfnamefont {J.~E.}\ \bibnamefont {Bowers}},\ }\bibfield  {title} {\bibinfo {title} {{3D integration enables
  ultralow-noise isolator-free lasers in silicon photonics}},\ }\href {https://doi.org/10.1038/s41586-023-06251-w} {\bibfield  {journal} {\bibinfo  {journal} {Nature}\ }\textbf {\bibinfo {volume} {620}},\ \bibinfo {pages} {78} (\bibinfo {year} {2023})}\BibitemShut {NoStop}%
\bibitem [{\citenamefont {Singleton}\ \emph {et~al.}(2018)\citenamefont {Singleton}, \citenamefont {Jouy}, \citenamefont {Beck},\ and\ \citenamefont {Faist}}]{Singleton2018EvidenceLasers}%
  \BibitemOpen
  \bibfield  {author} {\bibinfo {author} {\bibfnamefont {M.}~\bibnamefont {Singleton}}, \bibinfo {author} {\bibfnamefont {P.}~\bibnamefont {Jouy}}, \bibinfo {author} {\bibfnamefont {M.}~\bibnamefont {Beck}},\ and\ \bibinfo {author} {\bibfnamefont {J.}~\bibnamefont {Faist}},\ }\bibfield  {title} {\bibinfo {title} {{Evidence of linear chirp in mid-infrared quantum cascade lasers}},\ }\href {https://doi.org/10.1364/OPTICA.5.000948} {\bibfield  {journal} {\bibinfo  {journal} {Optica}\ }\textbf {\bibinfo {volume} {5}},\ \bibinfo {pages} {948} (\bibinfo {year} {2018})}\BibitemShut {NoStop}%
\bibitem [{\citenamefont {Opa{\v{c}}ak}\ and\ \citenamefont {Schwarz}(2019)}]{Opacak2019TheoryNonlinearity}%
  \BibitemOpen
  \bibfield  {author} {\bibinfo {author} {\bibfnamefont {N.}~\bibnamefont {Opa{\v{c}}ak}}\ and\ \bibinfo {author} {\bibfnamefont {B.}~\bibnamefont {Schwarz}},\ }\bibfield  {title} {\bibinfo {title} {{Theory of Frequency-Modulated Combs in Lasers with Spatial Hole Burning, Dispersion, and Kerr Nonlinearity}},\ }\href {https://doi.org/10.1103/PHYSREVLETT.123.243902} {\bibfield  {journal} {\bibinfo  {journal} {Physical Review Letters}\ }\textbf {\bibinfo {volume} {123}},\ \bibinfo {pages} {243902} (\bibinfo {year} {2019})}\BibitemShut {NoStop}%
\bibitem [{\citenamefont {Burghoff}(2020)}]{Burghoff2020UnravelingTheory}%
  \BibitemOpen
  \bibfield  {author} {\bibinfo {author} {\bibfnamefont {D.}~\bibnamefont {Burghoff}},\ }\bibfield  {title} {\bibinfo {title} {{Unraveling the origin of frequency modulated combs using active cavity mean-field theory}},\ }\href {https://doi.org/10.1364/OPTICA.408917} {\bibfield  {journal} {\bibinfo  {journal} {Optica}\ }\textbf {\bibinfo {volume} {7}},\ \bibinfo {pages} {1781} (\bibinfo {year} {2020})}\BibitemShut {NoStop}%
\bibitem [{\citenamefont {T{\"{a}}schler}\ \emph {et~al.}(2023)\citenamefont {T{\"{a}}schler}, \citenamefont {Forrer}, \citenamefont {Bertrand}, \citenamefont {Kapsalidis}, \citenamefont {Beck},\ and\ \citenamefont {Faist}}]{Taschler2023AsynchronousCombs}%
  \BibitemOpen
  \bibfield  {author} {\bibinfo {author} {\bibfnamefont {P.}~\bibnamefont {T{\"{a}}schler}}, \bibinfo {author} {\bibfnamefont {A.}~\bibnamefont {Forrer}}, \bibinfo {author} {\bibfnamefont {M.}~\bibnamefont {Bertrand}}, \bibinfo {author} {\bibfnamefont {F.}~\bibnamefont {Kapsalidis}}, \bibinfo {author} {\bibfnamefont {M.}~\bibnamefont {Beck}},\ and\ \bibinfo {author} {\bibfnamefont {J.}~\bibnamefont {Faist}},\ }\bibfield  {title} {\bibinfo {title} {{Asynchronous Upconversion Sampling of Frequency Modulated Combs}},\ }\href {https://doi.org/10.1002/LPOR.202200590} {\bibfield  {journal} {\bibinfo  {journal} {Laser {\&} Photonics Reviews}\ ,\ \bibinfo {pages} {220063}} (\bibinfo {year} {2023})}\BibitemShut {NoStop}%
\bibitem [{\citenamefont {Meng}\ \emph {et~al.}(2021)\citenamefont {Meng}, \citenamefont {Singleton}, \citenamefont {Hillbrand}, \citenamefont {Francki{\'{e}}}, \citenamefont {Beck},\ and\ \citenamefont {Faist}}]{Meng2021DissipativeLasers}%
  \BibitemOpen
  \bibfield  {author} {\bibinfo {author} {\bibfnamefont {B.}~\bibnamefont {Meng}}, \bibinfo {author} {\bibfnamefont {M.}~\bibnamefont {Singleton}}, \bibinfo {author} {\bibfnamefont {J.}~\bibnamefont {Hillbrand}}, \bibinfo {author} {\bibfnamefont {M.}~\bibnamefont {Francki{\'{e}}}}, \bibinfo {author} {\bibfnamefont {M.}~\bibnamefont {Beck}},\ and\ \bibinfo {author} {\bibfnamefont {J.}~\bibnamefont {Faist}},\ }\bibfield  {title} {\bibinfo {title} {{Dissipative Kerr solitons in semiconductor ring lasers}},\ }\href {https://doi.org/10.1038/s41566-021-00927-3} {\bibfield  {journal} {\bibinfo  {journal} {Nature Photonics}\ }\textbf {\bibinfo {volume} {16}},\ \bibinfo {pages} {142} (\bibinfo {year} {2021})}\BibitemShut {NoStop}%
\bibitem [{\citenamefont {Opa{\v{c}}ak}\ \emph {et~al.}(2023)\citenamefont {Opa{\v{c}}ak}, \citenamefont {Kazakov}, \citenamefont {Columbo}, \citenamefont {Beiser}, \citenamefont {Letsou}, \citenamefont {Pilat}, \citenamefont {Brambilla}, \citenamefont {Prati}, \citenamefont {Piccardo}, \citenamefont {Capasso},\ and\ \citenamefont {Schwarz}}]{Opacak2023Nozaki-BekkiSolitons}%
  \BibitemOpen
  \bibfield  {author} {\bibinfo {author} {\bibfnamefont {N.}~\bibnamefont {Opa{\v{c}}ak}}, \bibinfo {author} {\bibfnamefont {D.}~\bibnamefont {Kazakov}}, \bibinfo {author} {\bibfnamefont {L.~L.}\ \bibnamefont {Columbo}}, \bibinfo {author} {\bibfnamefont {M.}~\bibnamefont {Beiser}}, \bibinfo {author} {\bibfnamefont {T.~P.}\ \bibnamefont {Letsou}}, \bibinfo {author} {\bibfnamefont {F.}~\bibnamefont {Pilat}}, \bibinfo {author} {\bibfnamefont {M.}~\bibnamefont {Brambilla}}, \bibinfo {author} {\bibfnamefont {F.}~\bibnamefont {Prati}}, \bibinfo {author} {\bibfnamefont {M.}~\bibnamefont {Piccardo}}, \bibinfo {author} {\bibfnamefont {F.}~\bibnamefont {Capasso}},\ and\ \bibinfo {author} {\bibfnamefont {B.}~\bibnamefont {Schwarz}},\ }\bibfield  {title} {\bibinfo {title} {{Nozaki-Bekki optical solitons}},\ }\href {https://arxiv.org/abs/2304.10796v1} {\bibfield  {journal} {\bibinfo  {journal} {arXiv:2304.10796v1}\ } (\bibinfo {year} {2023})}\BibitemShut {NoStop}%
\bibitem [{\citenamefont {Leo}\ \emph {et~al.}(2010)\citenamefont {Leo}, \citenamefont {Coen}, \citenamefont {Kockaert}, \citenamefont {Gorza}, \citenamefont {Emplit},\ and\ \citenamefont {Haelterman}}]{Leo2010TemporalBuffer}%
  \BibitemOpen
  \bibfield  {author} {\bibinfo {author} {\bibfnamefont {F.}~\bibnamefont {Leo}}, \bibinfo {author} {\bibfnamefont {S.}~\bibnamefont {Coen}}, \bibinfo {author} {\bibfnamefont {P.}~\bibnamefont {Kockaert}}, \bibinfo {author} {\bibfnamefont {S.~P.}\ \bibnamefont {Gorza}}, \bibinfo {author} {\bibfnamefont {P.}~\bibnamefont {Emplit}},\ and\ \bibinfo {author} {\bibfnamefont {M.}~\bibnamefont {Haelterman}},\ }\bibfield  {title} {\bibinfo {title} {{Temporal cavity solitons in one-dimensional Kerr media as bits in an all-optical buffer}},\ }\href {https://doi.org/10.1038/nphoton.2010.120} {\bibfield  {journal} {\bibinfo  {journal} {Nature Photonics}\ }\textbf {\bibinfo {volume} {4}},\ \bibinfo {pages} {471} (\bibinfo {year} {2010})}\BibitemShut {NoStop}%
\bibitem [{\citenamefont {Herr}\ \emph {et~al.}(2013)\citenamefont {Herr}, \citenamefont {Brasch}, \citenamefont {Jost}, \citenamefont {Wang}, \citenamefont {Kondratiev}, \citenamefont {Gorodetsky},\ and\ \citenamefont {Kippenberg}}]{Herr2013TemporalMicroresonators}%
  \BibitemOpen
  \bibfield  {author} {\bibinfo {author} {\bibfnamefont {T.}~\bibnamefont {Herr}}, \bibinfo {author} {\bibfnamefont {V.}~\bibnamefont {Brasch}}, \bibinfo {author} {\bibfnamefont {J.~D.}\ \bibnamefont {Jost}}, \bibinfo {author} {\bibfnamefont {C.~Y.}\ \bibnamefont {Wang}}, \bibinfo {author} {\bibfnamefont {N.~M.}\ \bibnamefont {Kondratiev}}, \bibinfo {author} {\bibfnamefont {M.~L.}\ \bibnamefont {Gorodetsky}},\ and\ \bibinfo {author} {\bibfnamefont {T.~J.}\ \bibnamefont {Kippenberg}},\ }\bibfield  {title} {\bibinfo {title} {{Temporal solitons in optical microresonators}},\ }\href {https://doi.org/10.1038/nphoton.2013.343} {\bibfield  {journal} {\bibinfo  {journal} {Nature Photonics}\ }\textbf {\bibinfo {volume} {8}},\ \bibinfo {pages} {145} (\bibinfo {year} {2013})}\BibitemShut {NoStop}%
\bibitem [{\citenamefont {Henry}(1982)}]{Henry1982TheoryLasers}%
  \BibitemOpen
  \bibfield  {author} {\bibinfo {author} {\bibfnamefont {C.~H.}\ \bibnamefont {Henry}},\ }\bibfield  {title} {\bibinfo {title} {{Theory of the Linewidth of Semiconductor Lasers}},\ }\href {https://doi.org/10.1109/JQE.1982.1071522} {\bibfield  {journal} {\bibinfo  {journal} {IEEE Journal of Quantum Electronics}\ }\textbf {\bibinfo {volume} {18}},\ \bibinfo {pages} {259} (\bibinfo {year} {1982})}\BibitemShut {NoStop}%
\bibitem [{\citenamefont {Opa{\v{c}}ak}\ \emph {et~al.}(2021{\natexlab{a}})\citenamefont {Opa{\v{c}}ak}, \citenamefont {Cin}, \citenamefont {Hillbrand},\ and\ \citenamefont {Schwarz}}]{Opacak2021FrequencyNonlinearity}%
  \BibitemOpen
  \bibfield  {author} {\bibinfo {author} {\bibfnamefont {N.}~\bibnamefont {Opa{\v{c}}ak}}, \bibinfo {author} {\bibfnamefont {S.~D.}\ \bibnamefont {Cin}}, \bibinfo {author} {\bibfnamefont {J.}~\bibnamefont {Hillbrand}},\ and\ \bibinfo {author} {\bibfnamefont {B.}~\bibnamefont {Schwarz}},\ }\bibfield  {title} {\bibinfo {title} {{Frequency Comb Generation by Bloch Gain Induced Giant Kerr Nonlinearity}},\ }\href {https://doi.org/10.1103/PHYSREVLETT.127.093902/FIGURES/4/MEDIUM} {\bibfield  {journal} {\bibinfo  {journal} {Physical Review Letters}\ }\textbf {\bibinfo {volume} {127}},\ \bibinfo {pages} {093902} (\bibinfo {year} {2021}{\natexlab{a}})}\BibitemShut {NoStop}%
\bibitem [{\citenamefont {Opa{\v{c}}ak}\ \emph {et~al.}(2021{\natexlab{b}})\citenamefont {Opa{\v{c}}ak}, \citenamefont {Pilat}, \citenamefont {Kazakov}, \citenamefont {Cin}, \citenamefont {Ramer}, \citenamefont {Lendl}, \citenamefont {Capasso},\ and\ \citenamefont {Schwarz}}]{Opacak2021SpectrallyComb}%
  \BibitemOpen
  \bibfield  {author} {\bibinfo {author} {\bibfnamefont {N.}~\bibnamefont {Opa{\v{c}}ak}}, \bibinfo {author} {\bibfnamefont {F.}~\bibnamefont {Pilat}}, \bibinfo {author} {\bibfnamefont {D.}~\bibnamefont {Kazakov}}, \bibinfo {author} {\bibfnamefont {S.~D.}\ \bibnamefont {Cin}}, \bibinfo {author} {\bibfnamefont {G.}~\bibnamefont {Ramer}}, \bibinfo {author} {\bibfnamefont {B.}~\bibnamefont {Lendl}}, \bibinfo {author} {\bibfnamefont {F.}~\bibnamefont {Capasso}},\ and\ \bibinfo {author} {\bibfnamefont {B.}~\bibnamefont {Schwarz}},\ }\bibfield  {title} {\bibinfo {title} {{Spectrally resolved linewidth enhancement factor of a semiconductor frequency comb}},\ }\href {https://doi.org/10.1364/OPTICA.428096} {\bibfield  {journal} {\bibinfo  {journal} {Optica}\ }\textbf {\bibinfo {volume} {8}},\ \bibinfo {pages} {1227} (\bibinfo {year} {2021}{\natexlab{b}})}\BibitemShut {NoStop}%
\bibitem [{\citenamefont {Khurgin}(2023)}]{Khurgin2023NonlinearTime}%
  \BibitemOpen
  \bibfield  {author} {\bibinfo {author} {\bibfnamefont {J.~B.}\ \bibnamefont {Khurgin}},\ }\bibfield  {title} {\bibinfo {title} {{Nonlinear optics from the viewpoint of interaction time}},\ }\href {https://doi.org/10.1038/s41566-023-01191-3} {\bibfield  {journal} {\bibinfo  {journal} {Nature Photonics}\ }\textbf {\bibinfo {volume} {17}},\ \bibinfo {pages} {545} (\bibinfo {year} {2023})}\BibitemShut {NoStop}%
\bibitem [{\citenamefont {Shimoda}(1986)}]{Shimoda1986IntroductionPhysics}%
  \BibitemOpen
  \bibfield  {author} {\bibinfo {author} {\bibfnamefont {K.}~\bibnamefont {Shimoda}},\ }\href {https://doi.org/10.1007/978-3-540-38954-5} {\emph {\bibinfo {title} {{Introduction to Laser Physics}}}},\ Springer Series in Optical Sciences\ (\bibinfo  {publisher} {Springer Berlin Heidelberg},\ \bibinfo {address} {Berlin, Heidelberg},\ \bibinfo {year} {1986})\BibitemShut {NoStop}%
\bibitem [{\citenamefont {Kazakov}\ \emph {et~al.}(2022)\citenamefont {Kazakov}, \citenamefont {Letsou}, \citenamefont {Beiser}, \citenamefont {Zhi}, \citenamefont {Opa{\v{c}}ak}, \citenamefont {Piccardo}, \citenamefont {Schwarz},\ and\ \citenamefont {Capasso}}]{Kazakov2022SemiconductorCouplers}%
  \BibitemOpen
  \bibfield  {author} {\bibinfo {author} {\bibfnamefont {D.}~\bibnamefont {Kazakov}}, \bibinfo {author} {\bibfnamefont {T.~P.}\ \bibnamefont {Letsou}}, \bibinfo {author} {\bibfnamefont {M.}~\bibnamefont {Beiser}}, \bibinfo {author} {\bibfnamefont {Y.}~\bibnamefont {Zhi}}, \bibinfo {author} {\bibfnamefont {N.}~\bibnamefont {Opa{\v{c}}ak}}, \bibinfo {author} {\bibfnamefont {M.}~\bibnamefont {Piccardo}}, \bibinfo {author} {\bibfnamefont {B.}~\bibnamefont {Schwarz}},\ and\ \bibinfo {author} {\bibfnamefont {F.}~\bibnamefont {Capasso}},\ }\bibfield  {title} {\bibinfo {title} {{Semiconductor ring laser frequency combs with active directional couplers}},\ }\bibfield  {journal} {\bibinfo  {journal} {arXiv:2206.03379v2}\ }\href {https://doi.org/10.48550/arxiv.2206.03379} {10.48550/arxiv.2206.03379} (\bibinfo {year} {2022})\BibitemShut {NoStop}%
\bibitem [{\citenamefont {Joshi}\ \emph {et~al.}(2016)\citenamefont {Joshi}, \citenamefont {Jang}, \citenamefont {Luke}, \citenamefont {Ji}, \citenamefont {Miller}, \citenamefont {Klenner}, \citenamefont {Okawachi}, \citenamefont {Lipson},\ and\ \citenamefont {Gaeta}}]{Joshi2016ThermallyMicroresonators}%
  \BibitemOpen
  \bibfield  {author} {\bibinfo {author} {\bibfnamefont {C.}~\bibnamefont {Joshi}}, \bibinfo {author} {\bibfnamefont {J.~K.}\ \bibnamefont {Jang}}, \bibinfo {author} {\bibfnamefont {K.}~\bibnamefont {Luke}}, \bibinfo {author} {\bibfnamefont {X.}~\bibnamefont {Ji}}, \bibinfo {author} {\bibfnamefont {S.~A.}\ \bibnamefont {Miller}}, \bibinfo {author} {\bibfnamefont {A.}~\bibnamefont {Klenner}}, \bibinfo {author} {\bibfnamefont {Y.}~\bibnamefont {Okawachi}}, \bibinfo {author} {\bibfnamefont {M.}~\bibnamefont {Lipson}},\ and\ \bibinfo {author} {\bibfnamefont {A.~L.}\ \bibnamefont {Gaeta}},\ }\bibfield  {title} {\bibinfo {title} {{Thermally controlled comb generation and soliton modelocking in microresonators}},\ }\href {https://doi.org/10.1364/OL.41.002565} {\bibfield  {journal} {\bibinfo  {journal} {Optics Letters}\ }\textbf {\bibinfo {volume} {41}},\ \bibinfo {pages} {2565} (\bibinfo {year} {2016})}\BibitemShut {NoStop}%
\bibitem [{\citenamefont {Englebert}\ \emph {et~al.}(2021)\citenamefont {Englebert}, \citenamefont {Mas~Arab{\'{i}}}, \citenamefont {Parra-Rivas}, \citenamefont {Gorza},\ and\ \citenamefont {Leo}}]{Englebert2021TemporalResonator}%
  \BibitemOpen
  \bibfield  {author} {\bibinfo {author} {\bibfnamefont {N.}~\bibnamefont {Englebert}}, \bibinfo {author} {\bibfnamefont {C.}~\bibnamefont {Mas~Arab{\'{i}}}}, \bibinfo {author} {\bibfnamefont {P.}~\bibnamefont {Parra-Rivas}}, \bibinfo {author} {\bibfnamefont {S.-P.}\ \bibnamefont {Gorza}},\ and\ \bibinfo {author} {\bibfnamefont {F.}~\bibnamefont {Leo}},\ }\bibfield  {title} {\bibinfo {title} {{Temporal solitons in a coherently driven active resonator}},\ }\href {https://doi.org/10.1038/s41566-021-00807-w} {\bibfield  {journal} {\bibinfo  {journal} {Nature Photonics}\ }\textbf {\bibinfo {volume} {15}},\ \bibinfo {pages} {536} (\bibinfo {year} {2021})}\BibitemShut {NoStop}%
\bibitem [{\citenamefont {Trocha}\ \emph {et~al.}(2018)\citenamefont {Trocha}, \citenamefont {Karpov}, \citenamefont {Ganin}, \citenamefont {Pfeiffer}, \citenamefont {Kordts}, \citenamefont {Wolf}, \citenamefont {Krockenberger}, \citenamefont {Marin-Palomo}, \citenamefont {Weimann}, \citenamefont {Randel}, \citenamefont {Freude}, \citenamefont {Kippenberg},\ and\ \citenamefont {Koos}}]{Trocha2018UltrafastCombs}%
  \BibitemOpen
  \bibfield  {author} {\bibinfo {author} {\bibfnamefont {P.}~\bibnamefont {Trocha}}, \bibinfo {author} {\bibfnamefont {M.}~\bibnamefont {Karpov}}, \bibinfo {author} {\bibfnamefont {D.}~\bibnamefont {Ganin}}, \bibinfo {author} {\bibfnamefont {M.~H.}\ \bibnamefont {Pfeiffer}}, \bibinfo {author} {\bibfnamefont {A.}~\bibnamefont {Kordts}}, \bibinfo {author} {\bibfnamefont {S.}~\bibnamefont {Wolf}}, \bibinfo {author} {\bibfnamefont {J.}~\bibnamefont {Krockenberger}}, \bibinfo {author} {\bibfnamefont {P.}~\bibnamefont {Marin-Palomo}}, \bibinfo {author} {\bibfnamefont {C.}~\bibnamefont {Weimann}}, \bibinfo {author} {\bibfnamefont {S.}~\bibnamefont {Randel}}, \bibinfo {author} {\bibfnamefont {W.}~\bibnamefont {Freude}}, \bibinfo {author} {\bibfnamefont {T.~J.}\ \bibnamefont {Kippenberg}},\ and\ \bibinfo {author} {\bibfnamefont {C.}~\bibnamefont {Koos}},\ }\bibfield  {title} {\bibinfo {title} {{Ultrafast optical ranging using microresonator soliton frequency combs}},\ }\href
  {https://doi.org/10.1126/SCIENCE.AAO3924/SUPPL{\_}FILE/AAO3924{\_}TROCHA{\_}SM.PDF} {\bibfield  {journal} {\bibinfo  {journal} {Science}\ }\textbf {\bibinfo {volume} {359}},\ \bibinfo {pages} {887} (\bibinfo {year} {2018})}\BibitemShut {NoStop}%
\bibitem [{\citenamefont {Wang}\ \emph {et~al.}(2016)\citenamefont {Wang}, \citenamefont {Jaramillo-Villegas}, \citenamefont {Xuan}, \citenamefont {Xue}, \citenamefont {Bao}, \citenamefont {Leaird}, \citenamefont {Qi},\ and\ \citenamefont {Weiner}}]{Wang2016IntracavityRegime}%
  \BibitemOpen
  \bibfield  {author} {\bibinfo {author} {\bibfnamefont {P.-H.}\ \bibnamefont {Wang}}, \bibinfo {author} {\bibfnamefont {J.~A.}\ \bibnamefont {Jaramillo-Villegas}}, \bibinfo {author} {\bibfnamefont {Y.}~\bibnamefont {Xuan}}, \bibinfo {author} {\bibfnamefont {X.}~\bibnamefont {Xue}}, \bibinfo {author} {\bibfnamefont {C.}~\bibnamefont {Bao}}, \bibinfo {author} {\bibfnamefont {D.~E.}\ \bibnamefont {Leaird}}, \bibinfo {author} {\bibfnamefont {M.}~\bibnamefont {Qi}},\ and\ \bibinfo {author} {\bibfnamefont {A.~M.}\ \bibnamefont {Weiner}},\ }\bibfield  {title} {\bibinfo {title} {{Intracavity characterization of micro-comb generation in the single-soliton regime}},\ }\href {https://doi.org/10.1364/OE.24.010890} {\bibfield  {journal} {\bibinfo  {journal} {Optics Express}\ }\textbf {\bibinfo {volume} {24}},\ \bibinfo {pages} {10890} (\bibinfo {year} {2016})}\BibitemShut {NoStop}%
\bibitem [{\citenamefont {Xiang}\ \emph {et~al.}(2021)\citenamefont {Xiang}, \citenamefont {Liu}, \citenamefont {Guo}, \citenamefont {Chang}, \citenamefont {Wang}, \citenamefont {Weng}, \citenamefont {Peters}, \citenamefont {Xie}, \citenamefont {Zhang}, \citenamefont {Riemensberger}, \citenamefont {Selvidge}, \citenamefont {Kippenberg},\ and\ \citenamefont {Bowers}}]{Xiang2021LaserSilicon}%
  \BibitemOpen
  \bibfield  {author} {\bibinfo {author} {\bibfnamefont {C.}~\bibnamefont {Xiang}}, \bibinfo {author} {\bibfnamefont {J.}~\bibnamefont {Liu}}, \bibinfo {author} {\bibfnamefont {J.}~\bibnamefont {Guo}}, \bibinfo {author} {\bibfnamefont {L.}~\bibnamefont {Chang}}, \bibinfo {author} {\bibfnamefont {R.~N.}\ \bibnamefont {Wang}}, \bibinfo {author} {\bibfnamefont {W.}~\bibnamefont {Weng}}, \bibinfo {author} {\bibfnamefont {J.}~\bibnamefont {Peters}}, \bibinfo {author} {\bibfnamefont {W.}~\bibnamefont {Xie}}, \bibinfo {author} {\bibfnamefont {Z.}~\bibnamefont {Zhang}}, \bibinfo {author} {\bibfnamefont {J.}~\bibnamefont {Riemensberger}}, \bibinfo {author} {\bibfnamefont {J.}~\bibnamefont {Selvidge}}, \bibinfo {author} {\bibfnamefont {T.~J.}\ \bibnamefont {Kippenberg}},\ and\ \bibinfo {author} {\bibfnamefont {J.~E.}\ \bibnamefont {Bowers}},\ }\bibfield  {title} {\bibinfo {title} {{Laser soliton microcombs heterogeneously integrated on silicon}},\ }\href
  {https://doi.org/10.1126/SCIENCE.ABH2076/SUPPL{\_}FILE/ABH2076-XIANG-SM.PDF} {\bibfield  {journal} {\bibinfo  {journal} {Science}\ }\textbf {\bibinfo {volume} {373}},\ \bibinfo {pages} {99} (\bibinfo {year} {2021})}\BibitemShut {NoStop}%
\bibitem [{\citenamefont {Br{\`{e}}s}\ \emph {et~al.}(2023)\citenamefont {Br{\`{e}}s}, \citenamefont {Della~Torre}, \citenamefont {Grassani}, \citenamefont {Brasch}, \citenamefont {Grillet},\ and\ \citenamefont {Monat}}]{Bres2023SupercontinuumPerspectives}%
  \BibitemOpen
  \bibfield  {author} {\bibinfo {author} {\bibfnamefont {C.~S.}\ \bibnamefont {Br{\`{e}}s}}, \bibinfo {author} {\bibfnamefont {A.}~\bibnamefont {Della~Torre}}, \bibinfo {author} {\bibfnamefont {D.}~\bibnamefont {Grassani}}, \bibinfo {author} {\bibfnamefont {V.}~\bibnamefont {Brasch}}, \bibinfo {author} {\bibfnamefont {C.}~\bibnamefont {Grillet}},\ and\ \bibinfo {author} {\bibfnamefont {C.}~\bibnamefont {Monat}},\ }\bibfield  {title} {\bibinfo {title} {{Supercontinuum in integrated photonics: Generation, applications, challenges, and perspectives}},\ }\href {https://doi.org/10.1515/NANOPH-2022-0749/ASSET/GRAPHIC/J{\_}NANOPH-2022-0749{\_}FIG{\_}011.JPG} {\bibfield  {journal} {\bibinfo  {journal} {Nanophotonics}\ }\textbf {\bibinfo {volume} {12}},\ \bibinfo {pages} {1199} (\bibinfo {year} {2023})}\BibitemShut {NoStop}%
\bibitem [{\citenamefont {Liu}\ \emph {et~al.}(2022)\citenamefont {Liu}, \citenamefont {Qiu}, \citenamefont {Ji}, \citenamefont {Lukashchuk}, \citenamefont {He}, \citenamefont {Riemensberger}, \citenamefont {Hafermann}, \citenamefont {Ning~Wang}, \citenamefont {Liu}, \citenamefont {Ronning},\ and\ \citenamefont {Kippenberg}}]{Liu2022AAmplifier}%
  \BibitemOpen
  \bibfield  {author} {\bibinfo {author} {\bibfnamefont {Y.}~\bibnamefont {Liu}}, \bibinfo {author} {\bibfnamefont {Z.}~\bibnamefont {Qiu}}, \bibinfo {author} {\bibfnamefont {X.}~\bibnamefont {Ji}}, \bibinfo {author} {\bibfnamefont {A.}~\bibnamefont {Lukashchuk}}, \bibinfo {author} {\bibfnamefont {J.}~\bibnamefont {He}}, \bibinfo {author} {\bibfnamefont {J.}~\bibnamefont {Riemensberger}}, \bibinfo {author} {\bibfnamefont {M.}~\bibnamefont {Hafermann}}, \bibinfo {author} {\bibfnamefont {R.}~\bibnamefont {Ning~Wang}}, \bibinfo {author} {\bibfnamefont {J.}~\bibnamefont {Liu}}, \bibinfo {author} {\bibfnamefont {C.}~\bibnamefont {Ronning}},\ and\ \bibinfo {author} {\bibfnamefont {T.~J.}\ \bibnamefont {Kippenberg}},\ }\bibfield  {title} {\bibinfo {title} {{A photonic integrated circuit-based erbium-doped amplifier}},\ }\href {https://doi.org/10.1126/SCIENCE.ABO2631/SUPPL{\_}FILE/SCIENCE.ABO2631{\_}SM.PDF} {\bibfield  {journal} {\bibinfo  {journal} {Science}\ }\textbf {\bibinfo {volume} {376}},\ \bibinfo {pages}
  {1309} (\bibinfo {year} {2022})}\BibitemShut {NoStop}%
\bibitem [{\citenamefont {Gaafar}\ \emph {et~al.}(2023)\citenamefont {Gaafar}, \citenamefont {Ludwig}, \citenamefont {Wang}, \citenamefont {Wildi}, \citenamefont {Voumard}, \citenamefont {Sinobad}, \citenamefont {Lorenzen}, \citenamefont {Francis}, \citenamefont {Zhang}, \citenamefont {Bi}, \citenamefont {DeľHaye}, \citenamefont {Geiselmann}, \citenamefont {Singh}, \citenamefont {K{\"{a}}rtner}, \citenamefont {Garcia-Blanco},\ and\ \citenamefont {Herr}}]{Gaafar2023FemtosecondChip}%
  \BibitemOpen
  \bibfield  {author} {\bibinfo {author} {\bibfnamefont {M.~A.}\ \bibnamefont {Gaafar}}, \bibinfo {author} {\bibfnamefont {M.}~\bibnamefont {Ludwig}}, \bibinfo {author} {\bibfnamefont {K.}~\bibnamefont {Wang}}, \bibinfo {author} {\bibfnamefont {T.}~\bibnamefont {Wildi}}, \bibinfo {author} {\bibfnamefont {T.}~\bibnamefont {Voumard}}, \bibinfo {author} {\bibfnamefont {M.}~\bibnamefont {Sinobad}}, \bibinfo {author} {\bibfnamefont {J.}~\bibnamefont {Lorenzen}}, \bibinfo {author} {\bibfnamefont {H.}~\bibnamefont {Francis}}, \bibinfo {author} {\bibfnamefont {S.}~\bibnamefont {Zhang}}, \bibinfo {author} {\bibfnamefont {T.}~\bibnamefont {Bi}}, \bibinfo {author} {\bibfnamefont {P.}~\bibnamefont {DeľHaye}}, \bibinfo {author} {\bibfnamefont {M.}~\bibnamefont {Geiselmann}}, \bibinfo {author} {\bibfnamefont {N.}~\bibnamefont {Singh}}, \bibinfo {author} {\bibfnamefont {F.~X.}\ \bibnamefont {K{\"{a}}rtner}}, \bibinfo {author} {\bibfnamefont {S.~M.}\ \bibnamefont {Garcia-Blanco}},\ and\ \bibinfo {author} {\bibfnamefont
  {T.}~\bibnamefont {Herr}},\ }\bibfield  {title} {\bibinfo {title} {{Femtosecond pulse amplification on a chip}},\ }\href {https://arxiv.org/abs/2311.04758v1} {\bibfield  {journal} {\bibinfo  {journal} {arXiv:2311.04758v1}\ } (\bibinfo {year} {2023})}\BibitemShut {NoStop}%
\bibitem [{\citenamefont {Jung}\ \emph {et~al.}(2019)\citenamefont {Jung}, \citenamefont {Palaferri}, \citenamefont {Zhang}, \citenamefont {Xie}, \citenamefont {Okuno}, \citenamefont {Pinzone}, \citenamefont {Lascola},\ and\ \citenamefont {Belkin}}]{Jung2019HomogeneousPlatform}%
  \BibitemOpen
  \bibfield  {author} {\bibinfo {author} {\bibfnamefont {S.}~\bibnamefont {Jung}}, \bibinfo {author} {\bibfnamefont {D.}~\bibnamefont {Palaferri}}, \bibinfo {author} {\bibfnamefont {K.}~\bibnamefont {Zhang}}, \bibinfo {author} {\bibfnamefont {F.}~\bibnamefont {Xie}}, \bibinfo {author} {\bibfnamefont {Y.}~\bibnamefont {Okuno}}, \bibinfo {author} {\bibfnamefont {C.}~\bibnamefont {Pinzone}}, \bibinfo {author} {\bibfnamefont {K.}~\bibnamefont {Lascola}},\ and\ \bibinfo {author} {\bibfnamefont {M.~A.}\ \bibnamefont {Belkin}},\ }\bibfield  {title} {\bibinfo {title} {{Homogeneous photonic integration of mid-infrared quantum cascade lasers with low-loss passive waveguides on an InP platform}},\ }\href {https://doi.org/10.1364/OPTICA.6.001023} {\bibfield  {journal} {\bibinfo  {journal} {Optica}\ }\textbf {\bibinfo {volume} {6}},\ \bibinfo {pages} {1023} (\bibinfo {year} {2019})}\BibitemShut {NoStop}%
\bibitem [{\citenamefont {Wang}\ \emph {et~al.}(2022)\citenamefont {Wang}, \citenamefont {T{\"{a}}schler}, \citenamefont {Wang}, \citenamefont {Gini}, \citenamefont {Beck},\ and\ \citenamefont {Faist}}]{Wang2022MonolithicWaveguides}%
  \BibitemOpen
  \bibfield  {author} {\bibinfo {author} {\bibfnamefont {R.}~\bibnamefont {Wang}}, \bibinfo {author} {\bibfnamefont {P.}~\bibnamefont {T{\"{a}}schler}}, \bibinfo {author} {\bibfnamefont {Z.}~\bibnamefont {Wang}}, \bibinfo {author} {\bibfnamefont {E.}~\bibnamefont {Gini}}, \bibinfo {author} {\bibfnamefont {M.}~\bibnamefont {Beck}},\ and\ \bibinfo {author} {\bibfnamefont {J.}~\bibnamefont {Faist}},\ }\bibfield  {title} {\bibinfo {title} {{Monolithic Integration of Mid-Infrared Quantum Cascade Lasers and Frequency Combs with Passive Waveguides}},\ }\href {https://doi.org/10.1021/ACSPHOTONICS.1C01767/SUPPL{\_}FILE/PH1C01767{\_}SI{\_}001.PDF} {\bibfield  {journal} {\bibinfo  {journal} {ACS Photonics}\ }\textbf {\bibinfo {volume} {9}},\ \bibinfo {pages} {426} (\bibinfo {year} {2022})}\BibitemShut {NoStop}%
\bibitem [{\citenamefont {Dely}\ \emph {et~al.}(2022)\citenamefont {Dely}, \citenamefont {Bonazzi}, \citenamefont {Spitz}, \citenamefont {Rodriguez}, \citenamefont {Gacemi}, \citenamefont {Todorov}, \citenamefont {Pantzas}, \citenamefont {Beaudoin}, \citenamefont {Sagnes}, \citenamefont {Li}, \citenamefont {Giles~Davies}, \citenamefont {Linfield}, \citenamefont {Grillot}, \citenamefont {Vasanelli},\ and\ \citenamefont {Sirtori}}]{Dely202210Optoelectronics}%
  \BibitemOpen
  \bibfield  {author} {\bibinfo {author} {\bibfnamefont {H.}~\bibnamefont {Dely}}, \bibinfo {author} {\bibfnamefont {T.}~\bibnamefont {Bonazzi}}, \bibinfo {author} {\bibfnamefont {O.}~\bibnamefont {Spitz}}, \bibinfo {author} {\bibfnamefont {E.}~\bibnamefont {Rodriguez}}, \bibinfo {author} {\bibfnamefont {D.}~\bibnamefont {Gacemi}}, \bibinfo {author} {\bibfnamefont {Y.}~\bibnamefont {Todorov}}, \bibinfo {author} {\bibfnamefont {K.}~\bibnamefont {Pantzas}}, \bibinfo {author} {\bibfnamefont {G.}~\bibnamefont {Beaudoin}}, \bibinfo {author} {\bibfnamefont {I.}~\bibnamefont {Sagnes}}, \bibinfo {author} {\bibfnamefont {L.}~\bibnamefont {Li}}, \bibinfo {author} {\bibfnamefont {A.}~\bibnamefont {Giles~Davies}}, \bibinfo {author} {\bibfnamefont {E.~H.}\ \bibnamefont {Linfield}}, \bibinfo {author} {\bibfnamefont {F.}~\bibnamefont {Grillot}}, \bibinfo {author} {\bibfnamefont {A.}~\bibnamefont {Vasanelli}},\ and\ \bibinfo {author} {\bibfnamefont {C.}~\bibnamefont {Sirtori}},\ }\bibfield  {title} {\bibinfo {title} {{10
  Gbit/s Free Space Data Transmission at 9 µm Wavelength With Unipolar Quantum Optoelectronics}},\ }\href {https://doi.org/10.1002/LPOR.202100414} {\bibfield  {journal} {\bibinfo  {journal} {Laser {\&} Photonics Reviews}\ }\textbf {\bibinfo {volume} {16}},\ \bibinfo {pages} {2100414} (\bibinfo {year} {2022})}\BibitemShut {NoStop}%
\bibitem [{\citenamefont {Villares}\ \emph {et~al.}(2015)\citenamefont {Villares}, \citenamefont {Wolf}, \citenamefont {Kazakov}, \citenamefont {S{\"{u}}ess}, \citenamefont {Hugi}, \citenamefont {Beck},\ and\ \citenamefont {Faist}}]{Villares2015On-chipCombs}%
  \BibitemOpen
  \bibfield  {author} {\bibinfo {author} {\bibfnamefont {G.}~\bibnamefont {Villares}}, \bibinfo {author} {\bibfnamefont {J.}~\bibnamefont {Wolf}}, \bibinfo {author} {\bibfnamefont {D.}~\bibnamefont {Kazakov}}, \bibinfo {author} {\bibfnamefont {M.~J.}\ \bibnamefont {S{\"{u}}ess}}, \bibinfo {author} {\bibfnamefont {A.}~\bibnamefont {Hugi}}, \bibinfo {author} {\bibfnamefont {M.}~\bibnamefont {Beck}},\ and\ \bibinfo {author} {\bibfnamefont {J.}~\bibnamefont {Faist}},\ }\bibfield  {title} {\bibinfo {title} {{On-chip dual-comb based on quantum cascade laser frequency combs}},\ }\href {https://doi.org/10.1063/1.4938213} {\bibfield  {journal} {\bibinfo  {journal} {Applied Physics Letters}\ }\textbf {\bibinfo {volume} {107}},\ \bibinfo {pages} {251104} (\bibinfo {year} {2015})}\BibitemShut {NoStop}%
\bibitem [{\citenamefont {Prati}\ \emph {et~al.}(2021)\citenamefont {Prati}, \citenamefont {Brambilla}, \citenamefont {Piccardo}, \citenamefont {Columbo}, \citenamefont {Silvestri}, \citenamefont {Gioannini}, \citenamefont {Gatti}, \citenamefont {Lugiato},\ and\ \citenamefont {Capasso}}]{Prati2021SolitonSignal}%
  \BibitemOpen
  \bibfield  {author} {\bibinfo {author} {\bibfnamefont {F.}~\bibnamefont {Prati}}, \bibinfo {author} {\bibfnamefont {M.}~\bibnamefont {Brambilla}}, \bibinfo {author} {\bibfnamefont {M.}~\bibnamefont {Piccardo}}, \bibinfo {author} {\bibfnamefont {L.~L.}\ \bibnamefont {Columbo}}, \bibinfo {author} {\bibfnamefont {C.}~\bibnamefont {Silvestri}}, \bibinfo {author} {\bibfnamefont {M.}~\bibnamefont {Gioannini}}, \bibinfo {author} {\bibfnamefont {A.}~\bibnamefont {Gatti}}, \bibinfo {author} {\bibfnamefont {L.~A.}\ \bibnamefont {Lugiato}},\ and\ \bibinfo {author} {\bibfnamefont {F.}~\bibnamefont {Capasso}},\ }\bibfield  {title} {\bibinfo {title} {{Soliton dynamics of ring quantum cascade lasers with injected signal}},\ }\href {https://doi.org/10.1515/NANOPH-2020-0409/MACHINEREADABLECITATION/RIS} {\bibfield  {journal} {\bibinfo  {journal} {Nanophotonics}\ }\textbf {\bibinfo {volume} {10}},\ \bibinfo {pages} {195} (\bibinfo {year} {2021})}\BibitemShut {NoStop}%
\bibitem [{\citenamefont {Cox}\ and\ \citenamefont {Matthews}(2002)}]{Cox2002ExponentialSystems}%
  \BibitemOpen
  \bibfield  {author} {\bibinfo {author} {\bibfnamefont {S.~M.}\ \bibnamefont {Cox}}\ and\ \bibinfo {author} {\bibfnamefont {P.~C.}\ \bibnamefont {Matthews}},\ }\bibfield  {title} {\bibinfo {title} {{Exponential Time Differencing for Stiff Systems}},\ }\href {https://doi.org/10.1006/JCPH.2002.6995} {\bibfield  {journal} {\bibinfo  {journal} {Journal of Computational Physics}\ }\textbf {\bibinfo {volume} {176}},\ \bibinfo {pages} {430} (\bibinfo {year} {2002})}\BibitemShut {NoStop}%
\bibitem [{\citenamefont {Burghoff}\ \emph {et~al.}(2020)\citenamefont {Burghoff}, \citenamefont {Ren},\ and\ \citenamefont {Han}}]{Burghoff2020SensitivitySpectroscopy}%
  \BibitemOpen
  \bibfield  {author} {\bibinfo {author} {\bibfnamefont {D.}~\bibnamefont {Burghoff}}, \bibinfo {author} {\bibfnamefont {D.}~\bibnamefont {Ren}},\ and\ \bibinfo {author} {\bibfnamefont {Z.}~\bibnamefont {Han}},\ }\bibfield  {title} {\bibinfo {title} {{Sensitivity of SWIFT spectroscopy}},\ }\href {https://doi.org/10.1364/OE.382243} {\bibfield  {journal} {\bibinfo  {journal} {Optics Express}\ }\textbf {\bibinfo {volume} {28}},\ \bibinfo {pages} {6002} (\bibinfo {year} {2020})}\BibitemShut {NoStop}%
\end{thebibliography}%

\clearpage

\end{document}